\documentclass[12pt]{article}
\usepackage{amsmath,amssymb,bm,epsfig,psfrag}
\usepackage[square,comma,sort&compress]{natbib}

\allowdisplaybreaks 
\begin{document}

\begin{titlepage}

\begin{flushright}
MZ-TH/09-28\\[2mm]
August~26, 2009
\end{flushright}

\vspace{0.5cm}
\begin{center}
\Large\bf
Two-loop divergences of massive scattering amplitudes in non-abelian gauge theories
\end{center}

\vspace{0.5cm}
\begin{center}
{\sc Andrea Ferroglia, Matthias Neubert, Ben D.~Pecjak, and Li Lin Yang}\\
\vspace{0.4cm}
{\sl Institut f\"ur Physik (THEP), Johannes Gutenberg-Universit\"at\\
D-55099 Mainz, Germany}
\end{center}

\vspace{0.2cm}
\begin{abstract}\noindent
The infrared divergences of QCD scattering amplitudes can be derived from an anomalous dimension $\bm{\Gamma}$, which is a matrix in color space and depends on the momenta and masses of the external partons. It has recently been shown that in cases where there are at least two massive partons involved in the scattering process, starting at two-loop order $\bm{\Gamma}$ receives contributions involving color and momentum correlations between three (and more) partons. The three-parton correlations can be described by two universal functions $F_1$ and $f_2$. In this paper these functions are calculated at two-loop order in closed analytic form and their properties are studied in detail. Both functions are found to be suppressed like ${\cal O}(m^4/s^2)$ in the limit of small parton masses, in accordance with mass factorization theorems proposed in the literature. On the other hand, both functions are ${\cal O}(1)$ and even diverge logarithmically near the threshold for pair production of two heavy particles. As an application, we calculate the infrared poles in the $q\bar q\to t\bar t$ and $gg\to t\bar t$ scattering amplitudes at two-loop order.
\end{abstract}
\vfil

\end{titlepage}
\tableofcontents

\section{Introduction}

In the past few years, much progress has been achieved in the understanding of the infrared (IR) singularities of massless scattering amplitudes in non-abelian gauge theories, which are characterized by an intricate interplay of soft and collinear dynamics. While factorization proofs guarantee the absence of IR divergences in inclusive observables \cite{Collins:1989gx}, in many cases large Sudakov logarithms remain after this cancellation. A detailed control over the structure of IR poles in the virtual corrections to scattering amplitudes is a prerequisite for the resummation of these logarithms beyond the leading order \cite{Sterman:1986aj,Catani:1989ne,Contopanagos:1996nh,Kidonakis:1997gm}. Catani was the first to predict the singularities of two-loop scattering amplitudes apart from the $1/\epsilon$ pole term \cite{Catani:1998bh}, whose general form was derived later in \cite{MertAybat:2006wq,MertAybat:2006mz}. An interesting alternative approach to the problem of IR singularities was developed in \cite{Sterman:2002qn}, where the authors exploited the factorization properties of scattering amplitudes along with IR evolution equations to recover Catani's result at two-loop order and relate the coefficient of the $1/\epsilon$ pole term to a soft anomalous-dimension matrix. Surprisingly, the explicit calculation of the two-loop anomalous-dimension matrix revealed that it has the same color and momentum structure as at one-loop order \cite{MertAybat:2006wq,MertAybat:2006mz}. 

In recent work \cite{Becher:2009cu}, it was shown that the IR singularities of on-shell amplitudes in massless QCD are in one-to-one correspondence to the ultraviolet (UV) poles of operator matrix elements in soft-collinear effective theory (SCET) \cite{Bauer:2000yr,Bauer:2001yt,Beneke:2002ph}. They can be subtracted by means of a multiplicative $\bm{Z}$ factor, whose structure is constrained by the renormalization group. It was argued that the simplicity of the corresponding anomalous-dimension matrix holds not only at one- and two-loop order, but may in fact be an exact result of perturbation theory. This possibility was raised independently in \cite{Gardi:2009qi}. A non-trivial test of this prediction at three-loop order was performed in \cite{Dixon:2009gx}. Detailed theoretical arguments supporting this conjecture were presented in \cite{Becher:2009qa}, where constraints derived from soft-collinear factorization, the non-abelian exponentiation theorem \cite{Gatheral:1983cz,Frenkel:1984pz}, and the behavior of scattering amplitudes in two-parton collinear limits \cite{Kosower:1999xi} were employed to show that the anomalous-dimension matrix retains its simple form to three-loop order, with the possible exception of a single structure multiplying a function of two linearly independent conformal cross ratios built out of the momenta of four partons, which would have to vanish in all collinear limits. It was also proved that the coefficient of the Sudakov logarithm in the anomalous dimension obeys Casimir scaling at least through four-loop order.

It is interesting and relevant for many physical applications to generalize these results to the case of massive partons. The IR singularities of one-loop scattering amplitudes containing massive partons were obtained some time ago in \cite{Catani:2000ef}, but until very recently little was known about higher-loop results. In the limit where the parton masses are small compared with the typical momentum transfer among the partons, mass logarithms can be predicted based on collinear factorization theorems \cite{Mitov:2006xs,Becher:2007cu}. This allows one to obtain massive amplitudes from massless ones with a minimal amount of calculational effort. This method has been verified for two-loop QED amplitudes \cite{Becher:2007cu} and QCD amplitudes describing top-quark pair production at hadron colliders \cite{Czakon:2007ej,Czakon:2007wk}. A major step toward solving the problem of finding the IR divergences of generic two-loop scattering processes with both massive and massless partons, without the restriction to the limit of small masses, has been taken independently in \cite{Mitov:2009sv} and \cite{Becher:2009kw}. Interestingly, one finds that the simplicity of the anomalous-dimension matrix does not persist at two-loop order in the massive case. Important constraints from soft-collinear factorization and two-parton collinear limits are lost, and only the non-abelian exponentiation theorem restricts the color structures appearing in the anomalous-dimension matrix. At two-loop order, two different types of three-parton color and momentum correlations appear, whose effects can be parameterized in terms of two universal, process-independent functions $F_1$ and $f_2$ defined in equation~(\ref{resu1}) below. Apart from some symmetry properties, the precise form of these functions was left unspecified in \cite{Mitov:2009sv,Becher:2009kw}. 

The goal of this paper is to calculate and study these functions at two-loop order. A brief account of our main findings was presented in \cite{Ferroglia:2009ep}. In the following section we review known facts about the structure of the anomalous-dimension matrix governing the IR poles of scattering amplitudes in non-abelian gauge theories. Our main new contribution, the calculation of the universal functions $F_1$ and $f_2$, is described in Section~\ref{sec:F1f2}, where we also analyze some properties of these functions such as their analytic structure and their behavior near threshold. Contrary to statements made in the literature, we find that $F_1$ and $f_2$ do {\em not\/} vanish in the limit where the velocities of two massive partons approach each other. We then discuss the particularly interesting limit in which the parton masses are small compared with the typical momentum transfers. We find that in this limit our results are compatible with factorization theorems proposed in \cite{Mitov:2006xs,Becher:2007cu}. As an application of our general results, we present in Section~\ref{sec:tt} the anomalous-dimension matrices and $\bm{Z}$ factors relevant for the parton scattering processes $q\bar q, gg\to t\bar t+\mbox{colorless particles}$, where we allow for the presence of e.g.\ electroweak gauge bosons or Higgs bosons in the final state. Using these results, we derive analytic formulae for all $1/\epsilon^n$ pole terms at two-loop order in the squared $q\bar q\to t\bar t$ and $gg\to t\bar t$ scattering amplitudes, some of which were so far unknown or only known in numerical form. We attach these results in the form of a computer program. In Section~\ref{sec:elqq} we study elastic quark-quark scattering in the forward limit $s, m^2\gg|t|$, in which our general expression for the anomalous-dimension matrix reduces to the cross anomalous dimension studied a long time ago in \cite{Korchemskaya:1994qp}. This serves as a non-trivial check of our calculation. Our main findings are summarized in the concluding Section~\ref{sec:concl}.

\section{General structure of the anomalous dimension}
\label{sec:ADs}

We denote by $|{\cal M}_n(\epsilon,\{\underline{p}\},\{\underline{m}\})\rangle$ a UV-renormalized, on-shell $n$-parton scattering amplitude with IR singularities regularized in $d=4-2\epsilon$ dimensions. Here $\{\underline{p}\}\equiv\{p_1,\dots,p_n\}$ and $\{\underline{m}\}\equiv\{m_1,\dots,m_n\}$ denote the momenta and masses of the external partons. The amplitude is a function of the Lorentz invariants $s_{ij}\equiv 2\sigma_{ij}\,p_i\cdot p_j+i0$ and $p_i^2=m_i^2$, where the sign factor $\sigma_{ij}=+1$ if the momenta $p_i$ and $p_j$ are both incoming or outgoing, and $\sigma_{ij}=-1$ otherwise. For massive partons ($m_i\ne 0$), we define 4-velocities $v_i=p_i/m_i$ with $v_i^2=1$ and $v_i^0\ge 1$. We further define the recoil variables $w_{ij}\equiv-\sigma_{ij}\,v_i\cdot v_j-i0$. We use the color-space formalism of \cite{Catani:1996jh,Catani:1996vz}, in which $n$-particle amplitudes are treated as $n$-dimensional vectors in color space. $\bm{T}_i$ is the color generator associated with the $i$-th parton and acts as an $SU(N)$ matrix on its color indices. Specifically, one assigns $(\bm{T}_i^a)_{\alpha\beta}=t_{\alpha\beta}^a$ for a final-state quark or an initial-state anti-quark, $(\bm{T}_i^a)_{\alpha\beta}=-t_{\beta\alpha}^a$ for a final-state anti-quark or an initial-state quark, and $(\bm{T}_i^a)_{bc}=-if^{abc}$ for a gluon. We also use the notation $\bm{T}_i\cdot\bm{T}_j\equiv \bm{T}_i^a\,\bm{T}_j^a$ summed over $a$. Generators associated with different particles trivially commute, while $\bm{T}_i^2=C_i$ is given in terms of the eigenvalue of the quadratic Casimir operator of the corresponding color representation, i.e., $C_q=C_{\bar q}=C_F=\frac{N^2-1}{2N}$ for quarks and $C_g=C_A=N$ for gluons. Below, we will label massive partons with capital indices ($I,J,\dots$) and massless ones with lower-case indices ($i,j,\dots$). Note that color conservation implies the relation
\begin{equation}\label{colorsum}
   \sum_i\,\bm{T}_i + \sum_I\,\bm{T}_I = 0
\end{equation}
when acting on color-singlet states. 

It was shown in \cite{Becher:2009cu,Becher:2009qa,Becher:2009kw} that the IR poles of such amplitudes can be removed by a multiplicative renormalization factor $\bm{Z}^{-1}(\epsilon,\{\underline{p}\},\{\underline{m}\},\mu)$, which acts as a matrix on the color indices of the partons. More precisely, we have
\begin{equation}\label{Mnfinite}
   \bm{Z}^{-1}(\epsilon,\{\underline{p}\},\{\underline{m}\},\mu)\,
   |{\cal M}_n(\epsilon,\{\underline{p}\},\{\underline{m}\})\rangle
   \big|_{\alpha_s^{\rm QCD}\to\xi\alpha_s} = \mbox{finite}
\end{equation}
for $\epsilon\to 0$. The quantity $\alpha_s$ denotes the strong coupling constant in the effective theory, which is obtained after integrating out the heavy partons \cite{Becher:2009kw}. It is related to the coupling constant $\alpha_s^{\rm QCD}$ of full QCD via the decoupling relation $\alpha_s^{\rm QCD}=\xi\alpha_s$. In the relation above, this matching relation must be applied to the scattering amplitude in full QCD, as indicated. To first order in $\alpha_s$, the matching factor appropriate for $n_h$ heavy-quark flavors in QCD reads \cite{Steinhauser:2002rq}
\begin{equation}\label{eq:decouple}
   \xi = 1 + \frac{\alpha_s}{3\pi}\,T_F\,\sum_{i=1}^{n_h} 
   \left[ e^{\epsilon\gamma_E}\,\Gamma(\epsilon) 
   \left( \frac{\mu^2}{m_i^2} \right)^\epsilon - \frac{1}{\epsilon} 
   \right] + {\cal O}(\alpha_s^2) ,
\end{equation}
with $T_F=\frac12$. Here $m_i$ denote the masses of the heavy quarks.  Note that, as an alternative to (\ref{Mnfinite}), one can convert the expression for the $\bm{Z}$ factor from the effective to the full theory by replacing $\alpha_s\to\xi^{-1}\,\alpha_s^{\rm QCD}$. We will make use of this possibility in Section~\ref{sec:tt} to predict the IR poles of the $q\bar q\to t\bar t$ and $gg\to t\bar t$ amplitudes in full QCD.

The relation
\begin{equation}\label{RGE}
   \bm{Z}^{-1}(\epsilon,\{\underline{p}\},\{\underline{m}\},\mu)\,
   \frac{d}{d\ln\mu}\,
   \bm{Z}(\epsilon,\{\underline{p}\},\{\underline{m}\},\mu) 
   = - \bm{\Gamma}(\{\underline{p}\},\{\underline{m}\},\mu)
\end{equation}
links the renormalization factor to a universal anomalous-dimension matrix $\bm{\Gamma}$, which governs the scale dependence of effective-theory operators built out of collinear SCET fields for the massless partons and heavy-quark effective theory (HQET \cite{Neubert:1993mb}) fields for the massive ones. For the case of massless partons, the anomalous dimension has been calculated at two-loop order in \cite{MertAybat:2006wq,MertAybat:2006mz} and was found to contain only two-parton color-dipole correlations. It has recently been conjectured that this result may hold to all orders of perturbation theory \cite{Becher:2009cu,Gardi:2009qi,Becher:2009qa}. On the other hand, when massive partons are involved in the scattering process, then starting at two-loop order correlations involving more than two partons appear \cite{Mitov:2009sv}, the reason being that constraints from soft-collinear factorization and two-parton collinear limits, which protect the anomalous dimension in the massless case, no longer apply \cite{Becher:2009kw}. 

At two-loop order, the general structure of the anomalous-dimension matrix is \cite{Becher:2009kw}
\begin{eqnarray}\label{resu1}
   \bm{\Gamma}(\{\underline{p}\},\{\underline{m}\},\mu)
   &=& \sum_{(i,j)}\,\frac{\bm{T}_i\cdot\bm{T}_j}{2}\,
    \gamma_{\rm cusp}(\alpha_s)\,\ln\frac{\mu^2}{-s_{ij}}
    + \sum_i\,\gamma^i(\alpha_s) \nonumber\\
   &&\mbox{}- \sum_{(I,J)}\,\frac{\bm{T}_I\cdot\bm{T}_J}{2}\,
    \gamma_{\rm cusp}(\beta_{IJ},\alpha_s)
    + \sum_I\,\gamma^I(\alpha_s) 
    + \sum_{I,j}\,\bm{T}_I\cdot\bm{T}_j\,
    \gamma_{\rm cusp}(\alpha_s)\,\ln\frac{m_I\mu}{-s_{Ij}} 
    \nonumber\\
   &&\mbox{}+ \sum_{(I,J,K)} if^{abc}\, 
    \bm{T}_I^a\,\bm{T}_J^b\,\bm{T}_K^c\,
    F_1(\beta_{IJ},\beta_{JK},\beta_{KI}) \\
   &&\mbox{}+ \sum_{(I,J)} \sum_k\,if^{abc}\,
    \bm{T}_I^a\,\bm{T}_J^b\,\bm{T}_k^c\,
    f_2\Big(\beta_{IJ},
    \ln\frac{-\sigma_{Jk}\,v_J\cdot p_k}{-\sigma_{Ik}\,v_I\cdot p_k}
    \Big) + {\cal O}(\alpha_s^3) \,. \nonumber 
\end{eqnarray}
The one- and two-parton terms depicted in the first two lines start at one-loop order, while the three-parton terms in the last two lines start at ${\cal O}(\alpha_s^2)$. Starting at three-loop order also four-parton correlations would appear. The notation $(i,j,\dots)$ etc.\ refers to unordered tuples of distinct parton indices. We have defined the cusp angles $\beta_{IJ}$ via
\begin{equation}\label{wbrela}
   \cosh\beta_{IJ} = \frac{-s_{IJ}}{2m_I m_J} 
   = -\sigma_{IJ}\,v_I\cdot v_J-i0 = w_{IJ} \,.
\end{equation}
They are the hyperbolic angles formed by the time-like Wilson lines of two heavy partons. The physically allowed values for $w_{IJ}$ are $w_{IJ}\ge 1$ (one parton incoming and one outgoing), corresponding to $\beta_{IJ}\ge 0$, or $w_{IJ}\le -1$ (both partons incoming or outgoing), corresponding to $\beta_{IJ}=-b+i\pi$ with real $b\ge 0$.\footnote{This choice implies that $\sinh\beta=\sqrt{w^2-1}$. Alternatively, we could have used $\beta_{IJ}=b-i\pi$ with $b\ge 0$, in which case $\sinh\beta=w\,\sqrt{1-w^{-2}}$. We have confirmed that our results are the same in both cases.}
The first possibility corresponds to space-like kinematics, in which case the universal functions in (\ref{resu1}) are real. (For $f_2$ phases can arise if some of the variables $v_I\cdot p_k$ are time-like.) The second possibility corresponds to time-like kinematics, in which case these functions have non-trivial absorptive parts. For academic purposes it is sometimes useful to consider the analytic continuation of (\ref{wbrela}) to the Euclidean region $-1<w_{IJ}<1$, for which $\beta_{IJ}=ib$ with $0<b<\pi$ is purely imaginary. The form factors are real in this region, which however does not correspond to a physical scattering process. Later in this work we will consider the large-recoil limit $|w_{IJ}|\gg 1$, which is obtained whenever $m_I m_J\ll|s_{IJ}|$. In this limit we have $\beta_{IJ}=\ln(2w_{IJ})$ up to ${\cal O}(1/w_{IJ}^2)$ power corrections. 

The physically allowed values of the kinematical variables $w_{IJ}$, $w_{JK}$, and $w_{KI}$ entering the function $F_1$ are not arbitrary. To see this, note that in the rest frame of particle $I$ (where $v_I^0=1$) we have
\begin{equation}
   v_J\cdot v_K = (v_I\cdot v_J)\,(v_I\cdot v_K)
   - \cos\delta \sqrt{(v_I\cdot v_J)^2-1} \sqrt{(v_I\cdot v_K)^2-1} \,,
\end{equation}
where $\delta$ is the angle between $\vec{v}_J$ and $\vec{v}_K$, and all the scalar products satisfy $v_i\cdot v_j\ge 1$ by definition. Requiring that $\cos\delta$ be in the range $[-1,1]$ leads to the condition $G(|w_{IJ}|,|w_{JK}|,|w_{KI}|)\ge 0$ with 
\begin{equation}
   G(x,y,z) = 1 + 2xyz - (x^2+y^2+z^2) \,.
\end{equation}
Since in a three-parton configuration there is always at least one pair of partons either incoming or outgoing, at least one of the $w_{IJ}$ variables must be below $-1$, and hence the function $F_1$ is expected to have a non-zero imaginary part. For $f_2$ there is no non-trivial constraint on the allowed values of the kinematical variables, but also in this case there is at least one time-like invariant, so a non-trivial imaginary part arises.

The anomalous-dimension coefficients $\gamma_{\rm cusp}(\alpha_s)$ and $\gamma^i(\alpha_s)$ (for $i=q,g$) in (\ref{resu1}) have been determined to three-loop order in \cite{Becher:2009qa} by considering the cases of the massless quark and gluon form factors. Of particular importance for our discussion is the cusp anomalous dimension for light-like Wilson loops, whose two-loop expression is \cite{Korchemsky:1987wg}
\begin{equation}\label{gcusp2}
   \gamma_{\rm cusp}(\alpha_s) 
   = \frac{\alpha_s}{\pi} + \left( \frac{\alpha_s}{\pi} \right)^2
    \left[ \left( \frac{67}{36} - \frac{\pi^2}{12} \right) C_A 
    - \frac{5}{9}\,T_F n_l \right] + {\cal O}(\alpha_s^3) \,,
\end{equation}
where $n_l$ is the number of massless quark flavors. Similarly, the coefficients $\gamma^I(\alpha_s)$ for massive quarks and color-octet partons such as gluinos have been extracted at two-loop order in \cite{Becher:2009kw} by analyzing the anomalous dimension of heavy-light currents in SCET. In addition, the velocity-dependent function $\gamma_{\rm cusp}(\beta,\alpha_s)$ in (\ref{resu1}) was derived from the known two-loop anomalous dimension of a current composed of two heavy quarks moving at different velocity \cite{Korchemsky:1987wg,Korchemsky:1991zp}. A recent reanalysis of this anomalous dimension has led to the compact form \cite{Kidonakis:2009ev}
\begin{eqnarray}\label{Kido}
   \gamma_{\rm cusp}(\beta,\alpha_s)
   &=& \gamma_{\rm cusp}(\alpha_s)\,\beta\coth\beta \nonumber\\
   &&\hspace{-22mm}\mbox{}+ \frac{C_A}{2} 
    \left( \frac{\alpha_s}{\pi} \right)^2 
    \Bigg\{ \frac{\pi^2}{6} + \zeta_3 + \beta^2 
    + \coth^2\beta \left[ \mbox{Li}_3(e^{-2\beta}) 
    + \beta\,\mbox{Li}_2(e^{-2\beta}) - \zeta_3 
    + \frac{\pi^2}{6}\,\beta + \frac{\beta^3}{3} \right] \\
   &&\quad\mbox{}+ \coth\beta \left[ 
    \mbox{Li}_2(e^{-2\beta}) - 2\beta\,\ln(1-e^{-2\beta}) 
    - \frac{\pi^2}{6}\,(1+\beta) - \beta^2 - \frac{\beta^3}{3} 
    \right] \Bigg\}
    + {\cal O}(\alpha_s^3) \,. \nonumber
\end{eqnarray}
In the limit of large cusp angle one finds
\begin{equation}\label{gcuspexp}
   \gamma_{\rm cusp}(\beta,\alpha_s)
   = \gamma_{\rm cusp}(\alpha_s)\,\beta + {\cal O}(e^{-2\beta})
   = \gamma_{\rm cusp}(\alpha_s)\,\ln(2w)
    + {\cal O}\Big( \frac{1}{w^2} \Big) \,.
\end{equation}

Our goal in the present work is to further explore the structure of the three-parton correlation terms in (\ref{resu1}), which are parameterized by two universal functions: $F_1(x,y,z)$, which describes correlations involving three massive partons, is totally anti-symmetric in its arguments, while $f_2(x,y)$, which parameterizes correlations between a pair of massive partons and one massless parton, is an odd function of its second argument. The main technical challenge is to calculate these functions analytically at two-loop order, which involves the evaluation of a complicated, non-planar two-loop graph. This is the topic of the next section.

\section{Calculation of $\bm{F_1}$ and $\bm{f_2}$}
\label{sec:F1f2}

Our strategy will be to first consider the function $F_1$. To do so, we calculate the two-loop anomalous-dimension matrix of the soft Wilson-line operator $\bm{O}_s=\bm{S}_{v_1}(0)\,\bm{S}_{v_2}(0)\,\bm{S}_{v_3}(0)$ without imposing color conservation. This is important, since for three partons in a color-singlet state, color conservation would imply that $f^{abc}\,\bm{T}_1^a\,\bm{T}_2^b\,\bm{T}_3^c=-f^{abc}\,\bm{T}_1^a\,\bm{T}_2^b\,(\bm{T}_1^c + \bm{T}_2^c)=0$. The operator $\bm{O}_s$ consists of three time-like Wilson lines
\begin{equation}
   \bm{S}_v(x) = {\bf P} \exp\left( - ig\int_0^\infty\!dt\,
    v\cdot A_s^a(x+tv)\,\bm{T}^a \right)
\end{equation}
along the directions of the 4-velocities of three massive partons. The anomalous dimension of this operator is given by
\begin{equation}\label{resu2}
\begin{split}
   \bm{\Gamma}_s(\{\underline{v}\})
   &= \sum_{I=1}^3\,\gamma^I(\alpha_s) 
    - \Big[ \bm{T}_1\cdot\bm{T}_2\,
    \gamma_{\rm cusp}(\beta_{12},\alpha_s) 
    + \mbox{permutations} \Big] \\
   &\quad\mbox{}+ 6if^{abc}\,\bm{T}_1^a\,\bm{T}_2^b\,\bm{T}_3^c\,
    F_1(\beta_{12},\beta_{23},\beta_{31}) + {\cal O}(\alpha_s^3) \,. 
\end{split}
\end{equation}
The function $F_1$ follows from the coefficient of the $1/\epsilon$ pole in the bare matrix element of the operator. We will then obtain $f_2$ from a limiting procedure.

\subsection{Color-space formalism for Wilson coefficients and operators}

At this point a comment is in order concerning the color-space representation of on-shell scattering amplitudes in full QCD, and of Wilson coefficients and operators defined in the low-energy effective theory, which as mentioned earlier is a combination of SCET and HQET. In the effective theory, $n$-jet processes are described in terms of an effective Hamiltonian \cite{Becher:2009qa}
\begin{equation}\label{Heff}
   |{\cal H}_n^{\rm eff}(\{\underline{p}\},\{\underline{m}\})\rangle 
   = \sum\,\bm{O}_n(\{\underline{p}\},\{\underline{m}\},\mu)\,
    |{\cal C}_n(\{\underline{p}\},\{\underline{m}\},\mu)\rangle \,,
\end{equation}
where the short-distance Wilson coefficients are vectors in color space, whereas the operators $\bm{O}_n$ act as matrices on the color indices. The sum extends over all operators and coefficients that are color conserving in the sense of relation (\ref{colorsum}). The sum also extends over operators with different Dirac structure, but this point is irrelevant to our discussion here. Taking the inner product of $|{\cal H}_n^{\rm eff}\rangle$ with a given color state $\langle c\,|$ produces the color-stripped amplitude $\langle c\,|{\cal H}_n^{\rm eff}\rangle$, which is a $c$-number in color space.

The renormalized operators $\bm{O}_n(\mu)$ are related to the bare operators via a renormalization factor $\bm{Z}$, so that
\begin{equation}
   \bm{O}_n(\{\underline{p}\},\{\underline{m}\},\mu) 
   = \bm{O}_n^{\rm bare}(\epsilon,\{\underline{p}\},\{\underline{m}\})\,
    \bm{Z}(\epsilon,\{\underline{p}\},\{\underline{m}\},\mu) \,.
\end{equation}
The definition of the anomalous-dimension matrix in (\ref{RGE}) then implies the renormalization-group equation
\begin{equation}
  \label{eq:rgeo}
  \frac{d}{d\ln\mu}\,\bm{O}_n(\{\underline{p}\},\{\underline{m}\},\mu)
  = - \bm{O}_n(\{\underline{p}\},\{\underline{m}\},\mu)\,
  \bm{\Gamma}_n(\{\underline{p}\},\{\underline{m}\},\mu) \,.
\end{equation}
Similarly, from the fact that the effective Hamiltonian in (\ref{Heff}) is scale independent it follows that
\begin{equation}\label{Cnevol}
   \frac{d}{d\ln\mu}\,
    |{\cal C}_n(\{\underline{p}\},\{\underline{m}\},\mu)\rangle
   = \bm{\Gamma}_n(\{\underline{p}\},\{\underline{m}\},\mu)\,
    |{\cal C}_n(\{\underline{p}\},\{\underline{m}\},\mu)\rangle \,.
\end{equation}

As explained in \cite{Becher:2009cu,Becher:2009qa}, the parton scattering amplitudes in full QCD are given by the on-shell parton matrix elements of the effective Hamiltonian. Denoting these matrix elements by double brackets $\langle\!\langle\dots\rangle\!\rangle$, and using that on-shell parton matrix elements of the bare operators $\bm{O}_n^{\rm bare}$ are scaleless in the effective theory and are thus given by their tree-level expressions, we obtain
\begin{equation}
   |{\cal M}_n(\epsilon,\{\underline{p}\},\{\underline{m}\})\rangle
   = \sum\,\langle\!\langle\bm{O}_n^{\rm tree}\rangle\!\rangle\,
    \bm{Z}(\epsilon,\{\underline{p}\},\{\underline{m}\},\mu)\,
    |{\cal C}_n(\{\underline{p}\},\{\underline{m}\},\mu)\rangle \,.
\end{equation}
The tree-level matrix elements are given in terms of products of on-shell spinors and polarization vectors and act as unit matrices in color space, $\langle\!\langle\bm{O}_n^{\rm tree}\rangle\!\rangle=\langle\!\langle O_n^{\rm tree}\rangle\!\rangle\,\bm{1}$. It thus follows that
\begin{equation}\label{Mfinite}
\begin{split}
   |{\cal M}_n(\{\underline{p}\},\{\underline{m}\},\mu)\rangle
   &\equiv \lim_{\epsilon\to 0}\, 
    \bm{Z}^{-1}(\epsilon,\{\underline{p}\},\{\underline{m}\},\mu)\,
    |{\cal M}_n(\epsilon,\{\underline{p}\},\{\underline{m}\})\rangle \\
   &= \sum\,\langle\!\langle O_n^{\rm tree}\rangle\!\rangle\,
    |{\cal C}_n(\{\underline{p}\},\{\underline{m}\},\mu)\rangle 
\end{split}
\end{equation}
is the finite, subtracted amplitude introduced in (\ref{Mnfinite}). This quantity obeys the same evolution equation (\ref{Cnevol}) as the Wilson coefficients.

\subsection{Calculation of $\bm{F_1}$}

\begin{figure}
\begin{center}
\includegraphics{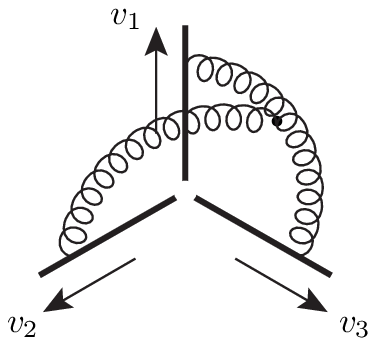}
\includegraphics{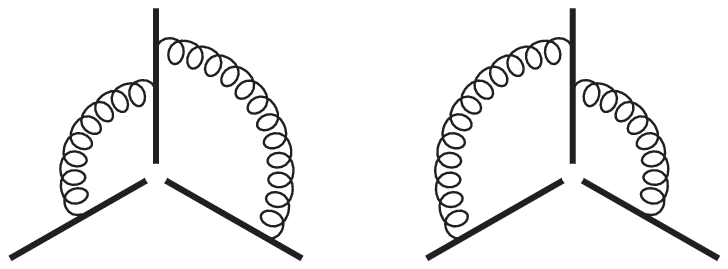}
\includegraphics{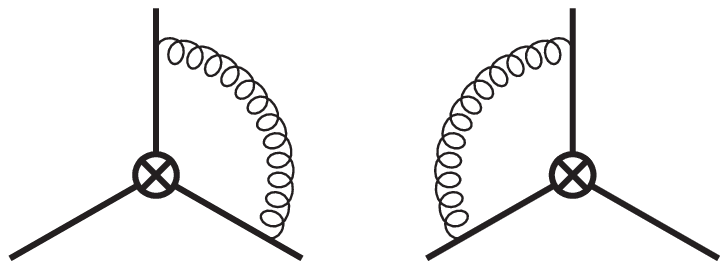}
\parbox{15cm}{\caption{\label{fig:dia} 
Two-loop Feynman graphs (top row) and one-loop counterterm diagrams (bottom row) contributing to the two-loop renormalization factor $\bm{Z}_s^{(2)}$.}}
\end{center}
\end{figure}

Consider now the vacuum expectation value of the soft operator $\bm{O}_s$ defined earlier. The renormalization factor $\bm{Z}_s$ for this operator is determined by the condition that $\langle\!\langle\bm{O}_s^{\rm bare}\rangle\!\rangle\,\bm{Z}_s$ be UV finite, and its anomalous dimension $\bm{\Gamma}_s$ is then determined from (\ref{RGE}). Here $\langle\!\langle\bm{O}_s^{\rm bare}\rangle\!\rangle$ denotes the bare vacuum matrix element calculated in HQET. Since now we are interested in the UV poles of these matrix elements, it is necessary to regularize IR singularities other than in dimensional regularization (see below). At two-loop order, the above renormalization condition gives
\begin{equation}
   \left[ \langle\!\langle\bm{O}_s^{\rm bare}\rangle\!\rangle^{(0)} 
   \bm{Z}_s^{(2)} 
   + \langle\!\langle\bm{O}_s^{\rm bare}\rangle\!\rangle^{(1)}
   \bm{Z}_s^{(1)}
   + \langle\!\langle\bm{O}_s^{\rm bare}\rangle\!\rangle^{(2)} 
   \right]_{\rm UV\,poles} = 0 \,,
\end{equation}
where here and below the superscripts in parenthesis refer in an obvious way to the order in the expansion in powers of $\alpha_s/4\pi$. The tree-level matrix element is $\langle\!\langle\bm{O}_s\rangle\!\rangle^{(0)}=1$. The equation above thus expresses the two-loop renormalization factor $\bm{Z}_s^{(2)}$ in terms of two contributions, 
\begin{equation}
   \bm{Z}_s^{(2)} 
   = - \left[ \langle\!\langle\bm{O}_s^{\rm bare}\rangle\!\rangle^{(2)} 
    + \langle\!\langle\bm{O}_s^{\rm bare}\rangle\!\rangle^{(1)} 
    \bm{Z}_s^{(1)} \right]_{\rm UV\,poles} .
\end{equation}
The function $F_1$ is derived from the pole terms in $\bm{Z}_s^{(2)}$ with totally anti-symmetric color structure, so we can limit the discussion to Feynman graphs involving the color generators of all three partons. Diagrammatically, the first contribution on the right-hand side contains the UV poles of the planar and non-planar two-loop graphs shown in the first row in Figure~\ref{fig:dia}. The second contribution corresponds to the UV poles of the one-loop diagrams with a counterterm insertion, as illustrated in the second row of the figure. In the calculation of the UV poles we regularize IR divergences by assigning residual external momenta $l_i$ to the Wilson lines, with $\omega_i\equiv-v_i\cdot l_i>0$. While the individual contributions depend on the $\omega_i$ regulators, their sum does not. Also, for concreteness we perform the calculation with three outgoing Wilson lines in the fundamental representation. Afterwards we replace the color matrices arising from the Feynman rules by $t^a\to\bm{T}^a$ to convert to the color-space formalism. For an incoming line the color matrix would get transposed, and in addition one would pick up a minus sign since the velocity in the corresponding heavy-quark propagator is reversed. As a result, in this case the correspondence would be $(-t^a)^T\to\bm{T}^a$, in accordance with the rules given in \cite{Catani:1996jh,Catani:1996vz}.

We find that the $1/\epsilon$ pole terms in the sum of all diagrams can be written as
\begin{eqnarray}
   \langle\!\langle\bm{O}_s^{\rm bare}\rangle\!\rangle^{(2)} 
    + \langle\!\langle\bm{O}_s^{\rm bare}\rangle\!\rangle^{(1)} 
    \bm{Z}_s^{(1)}
   &=& - \frac{2}{\epsilon^2} 
    \sum_{(I,J,K)} (\bm{T}_I\cdot\bm{T}_J)\,(\bm{T}_I\cdot\bm{T}_K)\,
    (\beta_{IJ}\,\coth\beta_{IJ})\,(\beta_{IK}\,\coth\beta_{IK}) 
    \nonumber\\
   &&\mbox{}- \frac{3}{2\epsilon}\, 
    if^{abc}\,\bm{T}_1^a\,\bm{T}_2^b\,\bm{T}_3^c\,
    F_1^{(2)}(\beta_{12},\beta_{23},\beta_{31}) + \dots \,,
\end{eqnarray}
where the dots represent finite terms and terms involving less than three different color generators. The double pole is multiplied by a symmetric color structure and receives contributions from the two-loop planar and one-loop counterterm diagrams only. It is given by the square of the one-loop anomalous-dimension matrix, as required by renormalization-group invariance \cite{Becher:2009qa}. We emphasize that the two-loop coefficient $F_1^{(2)}$ not only receives a contribution from the non-planar, triple-gluon graph depicted in the first diagram in Figure~\ref{fig:dia}, but also from the planar two-gluon diagrams and one-loop counterterm graphs. In \cite{Mitov:2009sv} a formal argument is sketched which suggests that the sum of the two planar two-loop graphs shown in the first row in Figure~\ref{fig:dia} has a vanishing coefficient multiplying the anti-symmetric color structure due to a cancellation between two divergent integrals. We find by explicit calculation that such a cancellation does not take place. Instead, the contribution to $F_1$ from the planar and counterterm diagrams reads 
\begin{equation}\label{F1planar}
\begin{split}
   F_1^{\rm (2)\,planar+CT}
   &= \frac{4}{3}\,\sum_{I,J,K} \epsilon_{IJK}\,
    \beta_{KI}\,\coth\beta_{KI}\,\coth\beta_{IJ} \\
   &\quad\times
    \left[ \beta_{IJ}^2 + 2\beta_{IJ} \ln(1-e^{-2\beta_{IJ}}) 
    - \mbox{Li}_2(e^{-2\beta_{IJ}}) + \frac{\pi^2}{6} \right] ,
\end{split}
\end{equation}
where the indices $I,J,K$ run over permutations of $(1,2,3)$. Moreover, as we shall see in Section~\ref{sec:small_mass}, important cancellations take place when the planar and non-planar contributions are combined.

We next consider the diagram with the triple gluon vertex, whose color factor $if^{abc}\,\bm{T}_1^a\,\bm{T}_2^b\,\bm{T}_3^c$ is totally anti-symmetric. Apart from this factor, the contribution of this diagram to $F_1$ is given by the two-loop integral
\begin{equation}
\begin{split}
   & (g_s\mu^\epsilon)^4 \int\frac{d^dk_1}{(2\pi)^d} 
    \frac{d^dk_2}{(2\pi)^d} \frac{d^dk_3}{(2\pi)^d}\,
    (2\pi)^d\,\delta^{(d)}(k_1+k_2+k_3) \\
   &\hspace{1.25cm}\times
    \frac{\epsilon_{IJK}\,w_{IJ}\,v_K\cdot k_I}%
         {k_1^2\,k_2^2\,k_3^2\,(v_1\cdot k_1-\omega_1)
          (v_2\cdot k_2-\omega_2)(v_3\cdot k_3-\omega_3)} \\
   &= \frac{2}{\epsilon} \left( \frac{\alpha_s}{4\pi} \right)^2 
    \epsilon_{IJK}\,I(w_{IJ},w_{JK},w_{KI}) + \mbox{UV finite.}
\end{split}
\end{equation}
The non-planar contribution to $F_1$ is
\begin{equation}\label{F2nonplanar}
   F_1^{\rm (2)\,non-planar} = \frac{4}{3}\, 
   \sum_{I,J,K} \epsilon_{IJK}\,I(w_{IJ},w_{JK},w_{KI}) \,.
\end{equation}
Introducing Feynman parameters and carrying out the loop momentum integrals, we obtain 
\begin{eqnarray}\label{eq:I1TFP}
   I(w_{12},w_{23},w_{31}) 
   &=& (w_{23}\,w_{31}+w_{12}) \int[dx][dy]\,x_1 y_3\,
    (x_1 x_2+x_2 x_3+x_3 x_1)^{-1} \\
   &&\hspace{-25mm}\times \left[ x_1 (y_2^2 + y_3^2 + 2w_{23} y_2 y_3)
    + x_2 (y_3^2 + y_1^2 + 2w_{31} y_3 y_1) 
    + x_3 (y_1^2 + y_2^2 + 2w_{12} y_1 y_2) \right]^{-2} , \nonumber
\end{eqnarray}
where $[dx]\equiv dx_1\,dx_2\,dx_3\,\delta(1-x_1-x_2-x_3)$, and all integrals run from 0 to 1. This result can be recast into the five-fold Mellin-Barnes representation
\begin{eqnarray}\label{eq:I1TMB}
   I(w_{12},w_{23},w_{31}) 
   &=& 2(w_{23}\,w_{31}+w_{12})\,\frac{1}{(2\pi i)^5}
    \int_{-i\infty}^{+i\infty}\!\bigg[ \prod_{i=1}^5 dz_i \bigg]\,
    (2w_{23})^{2z_1-1} (2w_{31})^{2z_2-1} (2w_{12})^{2z_3} 
    \nonumber\\
   &&\hspace{-2.0cm}\mbox{}\times 
    \frac{\Gamma(1-2z_1)\,\Gamma(1-2z_2)}%
    {\Gamma(z_1+z_2+z_3+z_4+z_5)}\,\Gamma(-2z_3)\,
    \Gamma(-z_4)\,\Gamma(z_1+z_3)\,\Gamma(z_1+z_5)\,
    \Gamma(z_2-z_5)\,\Gamma(z_3+z_5) \nonumber\\[1mm]
   &&\hspace{-2.0cm}\mbox{}\times 
    \Gamma(z_1+z_2+z_4)\,\Gamma(z_2+z_3+z_4)\,\Gamma(z_2+z_4+z_5)\,
    \Gamma(1-z_2-z_4-z_5) \,.
\end{eqnarray}
Decomposing the $w_{IJ}$ variables in terms of exponentials of cusp angles, $w_{IJ}=\cosh\beta_{IJ}=(\alpha_{IJ}+\alpha_{IJ}^{-1})/2$ with $\alpha_{IJ}\equiv e^{\beta_{IJ}}$, we can convert the factors $(2w_{IJ})^{2z_K}$ into powers of $\alpha_{IJ}$ by introducing three more Mellin-Barnes parameters. By applying Barnes' Lemmas repeatedly, we can then reduce the representation (\ref{eq:I1TMB}) to a three-fold one:
\begin{equation}
\begin{split}
   I(w_{12},w_{23},w_{31}) 
   &= 2(w_{23}\,w_{31}+w_{12})\,\frac{1}{(2\pi i)^3}
    \int_{-i\infty}^{+i\infty}\!dz_1\,dz_2\,dz_3\, 
    \alpha_{12}^{-2z_3} \alpha_{23}^{-1-2z_1} \alpha_{31}^{-1-2z2} \\
   &\quad\times \Gamma(-z_1-z_3)\,\Gamma(1+z_1-z_3)\,
    \Gamma(-z_1+z_3)\,\Gamma(1+z_1+z_3) \\[1mm]
   &\quad\times \Gamma^2(-z_2-z_3)\,\Gamma^2(1+z_2-z_3)\,
    \Gamma^2(-z_2+z_3)\,\Gamma^2(1+z_2+z_3) \,.
\end{split}
\end{equation}
The remaining integrals can be performed by closing the contours and summing up the residues. The resulting expression for $I$ is rather complicated, but the totally anti-symmetrized sum needed in (\ref{F2nonplanar}) turns out to be amazingly simple:
\begin{equation}\label{F1nonplanar}
   F_1^{\rm (2)\,non-planar} 
   = - \frac{4}{3} \sum_{I,J,K} \epsilon_{IJK}\,
    \beta_{IJ}^2\,\beta_{KI}\,\coth\beta_{KI} \,.
\end{equation}
In dealing with the Mellin-Barnes representations we have used the program package MB \cite{Czakon:2005rk} and associated packages found on the MB Tools web page \cite{MBTools}. We have checked the answer for this diagram numerically using sector decomposition \cite{Smirnov:2008py}. We have also checked that for Euclidean velocities our result for the triple-gluon diagram agrees numerically with a position-space based integral representation derived in \cite{Mitov:2009sv}. Combining all contributions, we finally find
\begin{equation}\label{eq:F1}
   F_1^{(2)}(\beta_{12},\beta_{23},\beta_{31}) 
   = \frac{4}{3} \sum_{I,J,K} \epsilon_{IJK}\,
    g(\beta_{IJ})\,\beta_{KI}\,\coth\beta_{KI} \,,
\end{equation}
where we have introduced the function
\begin{equation}
   g(\beta) = \coth\beta \left[ \beta^2 
    + 2\beta\,\ln(1-e^{-2\beta}) - \mbox{Li}_2(e^{-2\beta}) 
    + \frac{\pi^2}{6} \right] - \beta^2 - \frac{\pi^2}{6} \,. 
\end{equation}
The constant term $-\pi^2/6$ has been added by hand, so that $g(\beta)$ vanishes for $\beta\to\infty$. Its effect cancels in the anti-symmetrized sum over terms in (\ref{eq:F1}).

\subsection{Derivation of $\bm{f_2}$}

While the three-parton contribution described by $F_1$ is interesting on general grounds, there are not many processes of phenomenological importance in which three massive, colored partons are produced in a high-energy collision. For instance, searches for heavy, colored superpartners at the LHC will most likely focus on the pair production of squarks and gluinos. Hence, the three-parton term proportional to the function $f_2$ in (\ref{resu1}) is of greater practical importance. This function can be obtained from the result (\ref{eq:F1}) by writing $w_{23}=-\sigma_{23}\,v_2\cdot p_3/m_3$, $w_{31}=-\sigma_{31}\,v_1\cdot p_3/m_3$ and taking the limit $m_3\to 0$ at fixed $v_I\cdot p_3$. In that way, we obtain
\begin{equation}\label{f2limit}
   f_2\Big( \beta_{12}, 
    \ln\frac{-\sigma_{23}\,v_2\cdot p_3}%
            {-\sigma_{31}\,v_1\cdot p_3} \Big) 
   = 3 \lim_{m_3\to 0} F_1(\beta_{12},\beta_{23},\beta_{31}) \,.
\end{equation}
Starting from the expression for $F_1^{(2)}$ given earlier, we immediately derive the two-loop coefficient
\begin{equation}
   f_2^{(2)}\Big( \beta_{12}, 
    \ln\frac{-\sigma_{23}\,v_2\cdot p_3}%
            {-\sigma_{13}\,v_1\cdot p_3} \Big) 
   = - 4 g(\beta_{12})\,
    \ln\frac{-\sigma_{23}\,v_2\cdot p_3}%
            {-\sigma_{13}\,v_1\cdot p_3} \,,
\end{equation}
where $g(\beta)$ has been defined above. We believe it is not an accident that the function $f_2$ is linear in its second argument, but that this feature persists to all orders of perturbation theory. The reason is that the logarithm 
\begin{equation}\label{logdecomp}
   \ln\frac{-\sigma_{23}\,v_2\cdot p_3}{-\sigma_{13}\,v_1\cdot p_3}
   \equiv \ln\frac{-2\sigma_{23}\,v_2\cdot p_3}{\mu}
   - \ln\frac{-2\sigma_{13}\,v_1\cdot p_3}{\mu}
\end{equation}
is really the difference of two divergent collinear logarithms, and in order for the scale dependence to cancel between terms depending on one of the two logarithms, the dependence should be single logarithmic. 

\subsection{Properties of the universal functions}
\label{sec:prop}

We finish this section by collecting some useful properties of the three-parton correlation functions. We first note that, at least to two-loop order, we can rewrite the above relations in the suggestive form
\begin{equation}\label{beauty}
\begin{split}
   F_1(\beta_{12},\beta_{23},\beta_{31}) 
   &= \frac{1}{3} \sum_{I,J,K} \epsilon_{IJK}\,
    \frac{\alpha_s}{4\pi}\,g(\beta_{IJ})\,
    \gamma_{\rm cusp}(\beta_{KI},\alpha_s) \,, \\
   f_2\Big( \beta_{12}, 
    \ln\frac{-\sigma_{23}\,v_2\cdot p_3}%
            {-\sigma_{13}\,v_1\cdot p_3} \Big) 
   &= - \frac{\alpha_s}{4\pi}\,g(\beta_{12})\,
    \gamma_{\rm cusp}(\alpha_s)\,
    \ln\frac{-\sigma_{23}\,v_2\cdot p_3}%
            {-\sigma_{13}\,v_1\cdot p_3} \,,
\end{split}
\end{equation}
where $\gamma_{\rm cusp}(\beta,\alpha_s)$ and $\gamma_{\rm cusp}(\alpha_s)$ are the cusp anomalous dimensions entering the two-parton terms in (\ref{resu1}), and at one-loop order
\begin{equation}
   \gamma_{\rm cusp}(\beta,\alpha_s) 
   = \gamma_{\rm cusp}(\alpha_s)\,r(\beta) \,, 
    \qquad \mbox{with} \quad
   r(\beta) = \beta\,\coth\beta \,,
\end{equation}
where $\gamma_{\rm cusp}(\alpha_s)$ has been given in (\ref{gcusp2}). Whether a factorization of the three-parton terms into a cusp anomalous dimension times a function of another cusp angle persists at higher orders of perturbation theory is an interesting open question. 

\begin{figure}
\begin{center}
\psfrag{x}[b]{$w$}
\psfrag{y}[b]{$r(\beta(w))$}
\includegraphics[height=5.7cm]{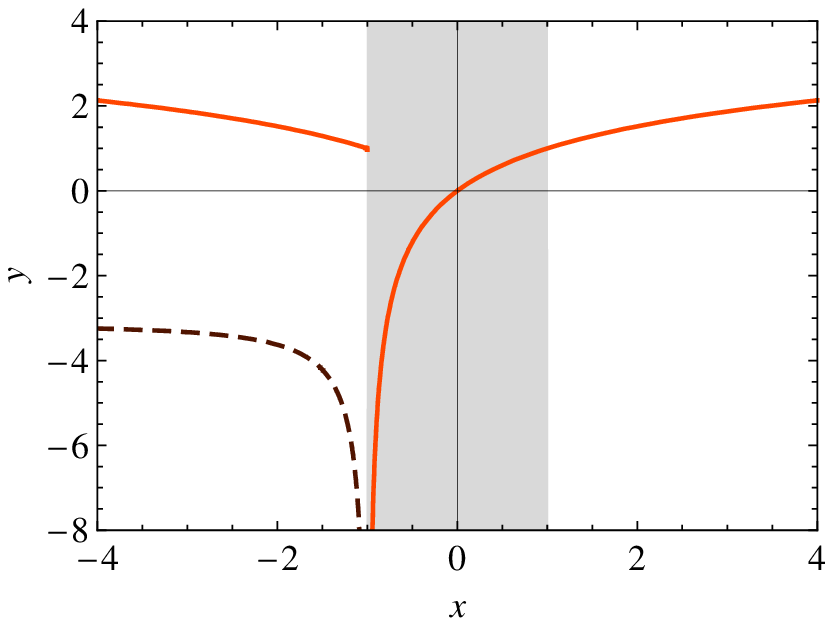}
\psfrag{x}[b]{$w$}
\psfrag{y}[b]{$g(\beta(w))$}
\quad
\includegraphics[height=5.7cm]{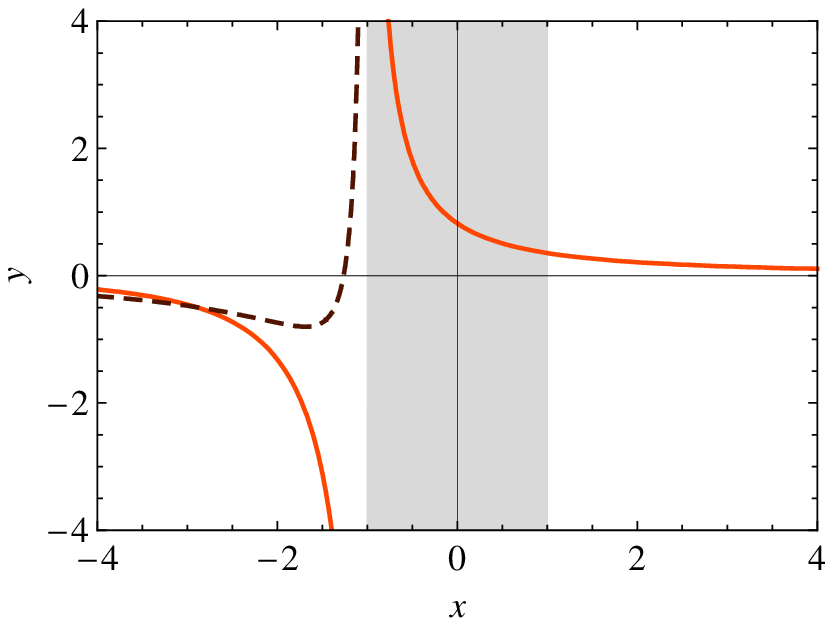}
\parbox{15cm}{\caption{\label{fig:rwgw} 
Graphical representation of the functions $r(\beta(w))$ and $g(\beta(w))$. Real parts are shown as solid lines, while imaginary parts, existing for $w<-1$, are shown as dashed lines. The regions $w\ge 1$ and $w\le -1$ correspond to space-like and time-like scattering, respectively. The shaded region $-1<w<1$ corresponds to Euclidean velocity vectors.}}
\end{center}
\end{figure}

It is useful to have expressions for the functions $F_1$ and $f_2$ in terms of the recoil variables $w_{IJ}=\cosh\beta_{IJ}$. These follow from 
\begin{eqnarray}
   r(\beta(w)) &=& \frac{w}{\sqrt{w^2-1}}\,
    \ln\Big(w+\sqrt{w^2-1}\Big) \,, \nonumber\\
   g(\beta(w)) &=& \frac{w}{\sqrt{w^2-1}} \left\{
    \ln\Big(w+\sqrt{w^2-1}\Big)\,
    \ln\frac{4(w^2-1)}{w+\sqrt{w^2-1}}
    - \mbox{Li}_2\Big[\Big(w-\sqrt{w^2-1}\Big)^{\!2}\,\Big] 
    + \frac{\pi^2}{6} \right\} \nonumber\\
   &&\mbox{}- \ln^2\!\Big(w+\sqrt{w^2-1}\Big) - \frac{\pi^2}{6} \,.
\end{eqnarray}
These functions are real for space-like $w\ge 1$, while for time-like $w\le -1$ they have discontinuities given by 
\begin{equation}
\begin{split}
   \frac{1}{\pi}\,\mbox{Im}\,r(\beta(w))
   &= \theta(-w-1)\,\frac{w}{\sqrt{w^2-1}} \,, \\
   \frac{1}{\pi}\,\mbox{Im}\,g(\beta(w)) 
   &= \theta(-w-1) \left[ 
    \frac{w}{\sqrt{w^2-1}}\,\ln\big[4(w^2-1)\big]
    + 2\ln\Big(-w+\sqrt{w^2-1}\Big) \right] .
\end{split}
\end{equation}
Here $w$ is always defined with imaginary part $-i0$, see (\ref{wbrela}). The function $f_2$ has further discontinuities in the logarithm given in (\ref{logdecomp}), provided that one of the scalar products $v_1\cdot p_3$ or $v_2\cdot p_3$ is time-like and the other one is space-like. In Figure~\ref{fig:rwgw} we show the real and imaginary parts of $r(\beta(w))$ and $g(\beta(w))$ as functions of $w=\cosh\beta$.

It is also interesting to expand about the zero-recoil point ($w=1$, $\beta=0$) and the threshold point ($w=-1$, $\beta=i\pi$). For the functions $r$ and $g$ near zero recoil, we find (with $w=1+\beta^2/2+\beta^4/24+\dots$, $\beta\ge 0$)
\begin{equation}\label{rgzero}
\begin{split}
   r(\beta) &= 1 + \frac{\beta^2}{3} - \frac{\beta^4}{45} 
    + {\cal O}(\beta^6) \,, \\
   g(\beta) &= \left( 2 - \frac{\pi^2}{6} \right) - \frac{\beta^2}{9}
    + \frac{14\beta^4}{675} + {\cal O}(\beta^6) \,.
\end{split}
\end{equation}
To expand the functions $F_1$ and $f_2$ themselves, we first note that at zero recoil $\beta_{12}=0$ and $\beta_{23}=i\pi-\beta_{31}$. Then, from (\ref{rgzero}) and the relations
\begin{equation}
\begin{aligned}
   r(i\pi-\beta) &= r(\beta) - i\pi\coth\beta \,, \\
   g(i\pi-\beta) &= g(\beta) - 2i\pi\coth\beta\,\ln(1-e^{-2\beta})
    - (\pi^2+2i\pi\beta)(\coth\beta-1) \,,
\end{aligned}
\end{equation}
we find that for $w_{12}\to 1$
\begin{equation}\label{eq:zrf1f2}
\begin{aligned}
   \lim_{\beta_{12}\to 0}\,F_1(\beta_{12},\beta_{23},\beta_{31}) 
   &= \frac{\alpha_s^2}{12\pi^2}\,\Big[ \pi^2 A(\beta_{31}) 
    + i\pi\,B(\beta_{31}) \Big] \,, \\
   \lim_{\beta_{12}\to 0}\,f_2\Big(\beta_{12}, 
    \ln\frac{-\sigma_{23}\,v_2\cdot p_3}{-\sigma_{13}\,v_1\cdot p_3}\Big)
   &= -\sigma_{13}\,\frac{\alpha_s^2}{4\pi^2}\,i\pi 
    \left( 2 - \frac{\pi^2}{6} \right) ,
\end{aligned}
\end{equation}
where 
\begin{equation}\label{eq:ab}
\begin{aligned}
   A(\beta) &= (\coth\beta-1)(\beta\coth\beta-1) \,, \\
   B(\beta) &= \coth^2\beta \left[ \mbox{Li}_2(e^{-2\beta})
    - \frac{\pi^2}{6} + \beta^2 \right] 
    + 2\coth\beta \left[ 1 - \beta - \frac{\beta^2}{2} 
    - \ln\left(1-e^{-2\beta}\right) \right] + 2\beta \,.
\end{aligned}
\end{equation}
We will make use of these results when studying elastic quark-quark scattering in Section~\ref{sec:elqq}.

Near threshold, the expansion of the functions $r$ and $g$ is (with $w=-1-b^2/2-b^4/24+\dots$, $b\ge 0$, and $\beta=-b+i\pi$)
\begin{equation}\label{rgthreshold}
\begin{aligned}
   r(\beta) &= - \frac{i\pi}{b} + 1 - \frac{i\pi b}{3} + \frac{b^2}{3} 
    + {\cal O}(b^3) \,, \\
   g(\beta) &= - \frac{\pi^2 + 2i\pi\ln(2b)}{b}
    + \left( 2 + \frac{5\pi^2}{6} \right) 
    - \frac{\pi^2 + 2i\pi\ln(2b) - 5i\pi}{3}\,b
    - \frac{b^2}{9} + {\cal O}(b^3) \,.
\end{aligned}
\end{equation}
Based on the symmetry properties of $F_1(\beta_{IJ},\beta_{JK},\beta_{KI})$ and $f_2(\beta_{IJ},\ln\frac{-\sigma_{Jk}\,v_J\cdot p_k}{-\sigma_{Ik}\,v_I\cdot p_k})$, it was concluded in \cite{Mitov:2009sv,Becher:2009kw} that these functions vanish whenever two of the 4-velocities of the massive partons coincide. Indeed, this seems to be an obvious consequence of the fact that $F_1$ is totally anti-symmetric in its arguments, while $f_2$ is odd in its second argument. This reasoning implicitly assumes that the limit of equal velocities is non-singular, but is invalidated by the presence of Coulomb singularities in $r(\beta)$ and $g(\beta)$ near threshold. In order to see this, consider the limit where two massive partons 1 and 2 are produced near threshold, i.e., $w_{12}\to -1$. In this case we can define the relative velocity $\vec{v}_{12}\equiv\vec{v}_1-\vec{v}_2$ in the frame where the two partons have equal and opposite velocities (for $m_1=m_2$ this is the rest frame of the pair), and $\vec{v}_{12}^2=4(w_{12}+1)/(w_{12}-1)=b^2+{\cal O}(b^4)$. To subleading order in the relative velocity, we thus have
\begin{equation}
   g(\beta_{12}) \big|_{\vec{v}_{12}\to 0} 
   = - \frac{\pi^2 + 2i\pi\ln(2|\vec{v}_{12}|)}{|\vec{v}_{12}|}
    + \left( 2 + \frac{5\pi^2}{6} \right) 
    + {\cal O}(|\vec{v}_{12}|) \,.
\end{equation}
It is then straightforward to derive the following limiting expressions valid near threshold (i.e., for $w_{12}\to -1$):
\begin{equation}
\begin{aligned}
   \lim_{\beta_{12}\to i\pi}\,F_1(\beta_{12},\beta_{23},\beta_{31}) 
   &= - \sigma_{13}\,\frac{\alpha_s^2}{12\pi^2}\,\Big\{ 
    \left[ \pi^2 + 2i\pi\ln(2|\vec{v}_{12}|) \right] r'(\beta_{31}) 
    - i\pi\,g'(\beta_{31}) \Big\} \\
   &\quad\times \left( \frac{\vec{v}_{12}}{|\vec{v}_{12}|}
    \cdot\frac{\vec{v}_3}{|\vec{v}_3|} \right)
    + {\cal O}(\alpha_s^3) \,, \\
   \lim_{\beta_{12}\to i\pi}\,f_2\Big(\beta_{12}, 
    \ln\frac{-\sigma_{23}\,v_2\cdot p_3}{-\sigma_{13}\,v_1\cdot p_3}\Big) 
   &= \frac{\alpha_s^2}{4\pi^2} \left[ \pi^2
    + 2i\pi\ln(2|\vec{v}_{12}|) \right]
    \frac{\vec{v}_{12}}{|\vec{v}_{12}|}
    \cdot\frac{\vec{p}_3}{|\vec{p}_3|} + {\cal O}(\alpha_s^3) \,.
\end{aligned}
\end{equation}
In both cases the scalar products of the unit vectors are nothing but $\cos\theta$, where $\theta$ is the scattering angle formed by the momenta of particles 1 and 3. In the first relation the primes denote derivatives with respect to $\beta_{31}$. The second relation is recovered from the first one using (\ref{f2limit}) and noting that $(-\sigma_{13})\,r'(\beta_{31})\to 1$ for $|\beta_{31}|\to\infty$. The above results are anti-symmetric in the parton indices 1 and 2 as required (note that $\beta_{13}=\beta_{23}$), but they do not vanish in the threshold limit. On the contrary, they diverge logarithmically in this limit.

\subsection{Limit of small parton masses}
\label{sec:small_mass}

A particularly interesting limit is that of large recoil, where all the scalar products $w_{IJ}$ become large in magnitude. According to the definition $w_{IJ}=-s_{IJ}/(2m_I m_J)$, this limit corresponds to $m_I m_J\to 0$ at fixed $s_{IJ}$. In this case, we find for the non-planar contribution in (\ref{F1nonplanar})
\begin{eqnarray}
   F_1^{\rm (2)\,non-planar} 
   &=& - \frac{4}{3}\,\ln\frac{w_{12}}{w_{23}}\,
    \ln\frac{w_{23}}{w_{31}}\,\ln\frac{w_{31}}{w_{12}} \\
   &&\mbox{}+ \sum_{I,J,K} \epsilon_{IJK}\,
    \frac{\ln(2w_{KI})}{3w_{IJ}^2}\,
    \Big[ 2\ln(2w_{IJ})\,\Big( \ln(2w_{KI}) + 1 \Big) - \ln(2w_{KI}) 
    \Big] + {\cal O}\Big( \frac{1}{w^3} \Big) \,. \nonumber
\end{eqnarray}
Similarly, the planar two-loop plus counterterm contributions in (\ref{F1planar}) can be expanded as
\begin{eqnarray}\label{F2planar}
   F_1^{\rm (2)\,planar+CT} 
   &=& \frac{4}{3}\,\ln\frac{w_{12}}{w_{23}}\,
    \ln\frac{w_{23}}{w_{31}}\,\ln\frac{w_{31}}{w_{12}} \nonumber\\
   &&\mbox{}+ \sum_{I,J,K} \epsilon_{IJK}\,
    \frac{\ln(2w_{KI})}{3w_{IJ}^2}\,
    \bigg[ 2\ln^2(2w_{IJ}) - 2\ln(2w_{IJ})\,\Big( \ln(2w_{KI}) 
    +2 \Big) \\
   &&\quad\mbox{}+ \ln(2w_{KI}) + \frac{\pi^2}{3} 
    - 1 \bigg] + {\cal O}\Big( \frac{1}{w^3} \Big) \,. \nonumber
\end{eqnarray}
Note that the leading terms in these expressions are unsuppressed, but they cancel in the sum of the two contributions. We then obtain
\begin{equation}\label{F1final}
   F_1^{(2)}(\beta_{12},\beta_{23},\beta_{31}) 
   = \sum_{I,J,K} \epsilon_{IJK}\,\frac{2\ln(2w_{KI})}{3w_{IJ}^2} 
    \left[ \ln^2(2w_{IJ}) - \ln(2w_{IJ}) + \frac{\pi^2}{6} 
    - \frac12 \right] + {\cal O}\Big( \frac{1}{w^3} \Big) \,,
\end{equation}
which can of course also be obtained directly from (\ref{eq:F1}). We see that $F_1$ vanishes in the limit $|w_{IJ}|\to\infty$. Similarly, in the limit of large recoil we have
\begin{equation}
   f_2^{(2)}\Big(\beta_{12},
    \ln\frac{-\sigma_{23}\,v_2\cdot p_3}%
            {-\sigma_{13}\,v_1\cdot p_3}\Big)
   = - \frac{2}{w_{12}^2} \left[ \ln^2(2w_{12}) - \ln(2w_{12}) 
    + \frac{\pi^2}{6} - \frac12 \right] 
    \ln\frac{-\sigma_{23}\,v_2\cdot p_3}{-\sigma_{13}\,v_1\cdot p_3}
    + {\cal O}\Big( \frac{1}{w_{12}^3} \Big) \,.
\end{equation}

It follows from these expressions that the three-parton correlation terms described by $F_1$ and $f_2$ vanish  like $(m_I m_J/s_{IJ})^2$ in the small-mass limit. This finding is in accordance with a factorization theorem proposed in \cite{Mitov:2006xs,Becher:2007cu}, which states that massive amplitudes in the small-mass limit can be obtained from massless ones by a simple rescaling prescription for the massive external legs. From (\ref{resu1}), (\ref{gcuspexp}), and (\ref{F1final}) we conclude that corrections to this simple factorization theorem are suppressed by {\em two\/} powers of $m_I m_J/s_{IJ}$. More specifically, we find that in the small-mass limit (see also the corresponding discussion for the two-parton terms in \cite{Becher:2009kw})
\begin{equation}
   \bm{\Gamma}(\{\underline{p}\},\{\underline{m}\},\mu)
   = \bm{\Gamma}(\{\underline{p}\},\mu) \\
    + \sum_I \left[ C_I\,\gamma_{\rm cusp}(\alpha_s)\,
    \ln\frac{\mu}{m_I} + \Delta\gamma^I(\alpha_s) \right] 
    + {\cal O}\Big(\frac{m_I^2 m_J^2}{s_{IJ}^2} \Big) \,,
\end{equation}
where $\bm{\Gamma}(\{\underline{p}\},\mu)$ is the anomalous-dimension matrix in the massless case, and $\Delta\gamma^Q(\alpha_s)\equiv\gamma^Q(\alpha_s)-\gamma^q(\alpha_s)$ (and similarly for hypothetical, heavy color-octet partons) is the difference of the single-parton anomalous dimensions belonging to massive and massless quarks. The extra terms give rise to one-particle $Z_I$ factors for each massive parton involved in the scattering process. As explained in \cite{Becher:2009kw}, in the limit of small parton masses the product $\prod_I Z_I^{-1}$ removes the IR poles in the ratio of the massive to massless amplitudes (without including loops of heavy partons\footnote{One can account for heavy-parton loops by applying the inverse of the decoupling relation (\ref{eq:decouple}) to the combined $\bm{Z}$ factor, see Section~\ref{subsec:IRpoles}.}). We emphasize that such a simple procedure will no longer work when ${\cal O}(m_I^2 m_J^2/s_{IJ}^2)$ corrections are included, since in this case novel IR poles with non-trivial color and momentum correlations arise.

\section{IR singularities in top-quark pair production}
\label{sec:tt}

As a first application, we employ our formalism to calculate the two-loop anomalous-dimension matrices for top-quark pair production in the $q\bar q\to t\bar t$ and $gg\to t\bar t$ channels and use them to construct the IR poles in the virtual corrections to these processes at two-loop order. These anomalous-dimension matrices form the basis for soft-gluon resummation at the next-to-next-to-leading logarithmic (NNLL) order for generic kinematics, i.e., not necessarily restricted to the threshold region.

\subsection{Anomalous-dimension matrices}

The first step is to derive the explicit form of the anomalous-dimension matrix (\ref{resu1}) in a given color basis for the partonic amplitudes (see, e.g., \cite{Kidonakis:1997gm,Kidonakis:1998nf}). We adopt the $s$-channel singlet-octet basis, in which the $t\bar t$ pair is either in a color-singlet or color-octet state. For the quark-antiquark annihilation process $q_l(p_1)+\bar q_k(p_2)\to t_i(p_3)+\bar t_j(p_4)$, we thus choose the independent color structures as
\begin{equation}\label{eq:qqbasis}
   |c_1\rangle = \delta_{ij}\,\delta_{kl} \,, 
    \qquad 
   |c_2\rangle = (t^a)_{ij}\,(t^a)_{kl} \,.
\end{equation}
For the gluon fusion process $g_a(p_1)+g_b(p_2)\to t_i(p_3)+\bar t_j(p_4)$, we use the basis
\begin{equation}
   |c_1\rangle = \delta^{ab}\,\delta_{ij} \,,
    \qquad 
   |c_2\rangle = if^{abc}\,(t^c)_{ij} \,,
    \qquad 
   |c_3\rangle = d^{abc}\,(t^c)_{ij} \,.
\end{equation}
Here $a$, $b$, $i$, $j$, $k$, $l$ are color indices. We find that the anomalous-dimension matrix for the $q\bar q$ channel can be written in the form 
\begin{equation}\label{eq:qqmatrix}
\begin{split}
   \bm{\Gamma}_{q\bar q} 
   &= \left[ C_F\,\gamma_{\rm cusp}(\alpha_s)\,\ln\frac{-s}{\mu^2}
    + C_F\,\gamma_{\rm cusp}(\beta_{34},\alpha_s)
    + 2\gamma^q(\alpha_s) + 2\gamma^Q(\alpha_s) \right] \bm{1} \\
   &\quad\mbox{}+ \frac{N}{2} \left[ 
    \gamma_{\rm cusp}(\alpha_s)\,
    \ln\frac{(-s_{13})(-s_{24})}{(-s)\,m_t^2}
    - \gamma_{\rm cusp}(\beta_{34},\alpha_s) \right]
    \begin{pmatrix}
     0~ & ~0 \\ 0~ & ~1
    \end{pmatrix} \\
   &\quad\mbox{}+ \gamma_{\rm cusp}(\alpha_s)\,
    \ln\frac{(-s_{13})(-s_{24})}{(-s_{14})(-s_{23})} \left[
    \begin{pmatrix}
     0~ & \frac{C_F}{2N} \\
     1~ & - \frac{1}{N}
    \end{pmatrix} 
    + \frac{\alpha_s}{4\pi}\,g(\beta_{34})
    \begin{pmatrix}
     0 & \frac{C_F}{2} \\ -N & 0
    \end{pmatrix} \right]
    + {\cal O}(\alpha_s^3) \,,
\end{split}
\end{equation}
where $s\equiv s_{12}$ is the square of the partonic center-of-mass energy. The term proportional to $g(\beta_{34})$ stems from the three-parton contributions
\begin{equation}
   - \left[ f_2\Big(\beta_{34},\ln\frac{-s_{13}}{-s_{14}}\Big) 
   + f_2\Big(\beta_{34},\ln\frac{-s_{24}}{-s_{23}}\Big) \right]
   \begin{pmatrix}
    0 & \frac{C_F}{2} \\ -N & 0
   \end{pmatrix} .
\end{equation}
With the help of the second relation in (\ref{beauty}) this can be recast into the product of $g(\beta_{34})$ times a conformal cross ratio \cite{Gardi:2009qi} of four momentum invariants. Similarly, for the $gg$ channel we obtain
\begin{eqnarray}
\label{eq:ggmatrix}
   \bm{\Gamma}_{gg} 
   &=& \left[ N\,\gamma_{\rm cusp}(\alpha_s)\,\ln\frac{-s}{\mu^2} 
    + C_F\,\gamma_{\rm cusp}(\beta_{34},\alpha_s)
    + 2\gamma^g(\alpha_s) + 2\gamma^Q(\alpha_s) \right] \bm{1} 
    \nonumber\\
   &&\mbox{}+ \frac{N}{2} \left[ 
    \gamma_{\rm cusp}(\alpha_s)\,
    \ln\frac{(-s_{13})(-s_{24})}{(-s)\,m_t^2}
    - \gamma_{\rm cusp}(\beta_{34},\alpha_s) \right]
    \begin{pmatrix}
     0~ & ~0~ & ~0 \\
     0~ & ~1~ & ~0 \\
     0~ & ~0~ & ~1
    \end{pmatrix} \\
   &&\mbox{}+ \gamma_{\rm cusp}(\alpha_s)\,
    \ln\frac{(-s_{13})(-s_{24})}{(-s_{14})(-s_{23})} \left[
    \begin{pmatrix}
     0~ & \frac{1}{2} & 0 \\
     1~ & - \frac{N}{4} & \frac{N^2-4}{4N} \\
     0~ & \frac{N}{4} & - \frac{N}{4}
    \end{pmatrix} 
    + \frac{\alpha_s}{4\pi}\,g(\beta_{34})
    \begin{pmatrix}
     0 & \frac{N}{2} & 0~ \\ -N & ~0~ & 0~ \\ 0 & ~0~ & 0~
    \end{pmatrix} \right] 
    + {\cal O}(\alpha_s^3) \,. \nonumber
\end{eqnarray}
When deriving these results we did not impose momentum conservation. Therefore, they can also be used for processes involving additional colorless particles such as electroweak gauge bosons or Higgs bosons.

The above anomalous-dimension matrices are the ingredients needed for soft-gluon resummation at NNLL order for general kinematics, a topic which we leave for future work. An interesting limit, which is often discussed in the literature, is the threshold limit $s\to 4m_t^2$. For this purpose it is convenient to define the quantity $\beta_t=\sqrt{1-4m_t^2/s}$, which is related to the relative 3-velocity $\vec{v}_{t\bar t}$ between the top-quark pair in the center-of-mass frame by $|\vec{v}_{t\bar t}|=2\beta_t$. The kinematical invariants can be expressed in terms of $\beta_t$ and the scattering angle $\theta$ between partons 1 and 3 in this frame according to
\begin{equation}
   \beta_{34} = i\pi - \ln\frac{1+\beta_t}{1-\beta_t} \,, 
    \quad 
   s_{13} = s_{24} = - \frac{s}{2}\,(1-\beta_t\cos\theta) \,,
    \quad
   s_{14} = s_{23} = - \frac{s}{2}\,(1+\beta_t\cos\theta) \,.
\end{equation}
Using the expansions in (\ref{rgthreshold}), we find that in the threshold limit $\beta_t\to 0$ the two-loop anomalous-dimension matrices reduce to
\begin{equation}\label{Gqqthresh}
\begin{split}
   \bm{\Gamma}_{q\bar q} 
   &= \left[ C_F\,\gamma_{\rm cusp}(\alpha_s)
    \left( \ln\frac{s}{\mu^2} - \frac{i\pi}{2\beta_t} - i\pi + 1
    \right)
    + C_F\,\gamma_{\rm cusp}^{(2)}(\beta_t)
    + 2\gamma^q(\alpha_s) + 2\gamma^Q(\alpha_s) \right] \bm{1} \\
   &\quad\mbox{}+ \frac{N}{2} \left[ \gamma_{\rm cusp}(\alpha_s)
    \left( \frac{i\pi}{2\beta_t} + i\pi - 1 \right)
    - \gamma_{\rm cusp}^{(2)}(\beta_t) \right]
    \begin{pmatrix}
     0~ & ~0 \\ 0~ & ~1
    \end{pmatrix} \\
   &\quad\mbox{}+ \frac{\alpha_s^2}{2\pi^2}
    \left[ \pi^2 + 2i\pi\ln(4\beta_t) \right] \cos\theta
    \begin{pmatrix}
     0 & \frac{C_F}{2} \\ -N & 0
    \end{pmatrix}
    + {\cal O}(\beta_t) + {\cal O}(\alpha_s^3) \,,
\end{split}
\end{equation}
and
\begin{eqnarray}\label{Gggthresh}
   \bm{\Gamma}_{gg} 
   &=& \left[ \gamma_{\rm cusp}(\alpha_s) 
    \left\{ N \left( \ln\frac{s}{\mu^2} - i\pi \right)
    - C_F \left( \frac{i\pi}{2\beta_t} - 1 \right) \right\}
    + C_F\,\gamma_{\rm cusp}^{(2)}(\beta_t)
    + 2\gamma^g(\alpha_s) + 2\gamma^Q(\alpha_s) \right] \bm{1} 
    \nonumber\\
   &&\mbox{}+ \frac{N}{2} \left[ \gamma_{\rm cusp}(\alpha_s)
    \left( \frac{i\pi}{2\beta_t} + i\pi - 1 \right)
    - \gamma_{\rm cusp}^{(2)}(\beta_t) \right]
    \begin{pmatrix}
     0~ & ~0~ & ~0 \\
     0~ & ~1~ & ~0 \\
     0~ & ~0~ & ~1
    \end{pmatrix} \\
   &&\mbox{}+ \frac{N\alpha_s^2}{2\pi^2}
    \left[ \pi^2 + 2i\pi\ln(4\beta_t) \right] \cos\theta
    \begin{pmatrix}
     0 & \frac12 & 0~ \\ -1 & ~0~ & 0~ \\ 0 & ~0~ & 0~
    \end{pmatrix} 
    + {\cal O}(\beta_t) + {\cal O}(\alpha_s^3) \,, \nonumber
\end{eqnarray}
where the two-loop expression for $\gamma_{\rm cusp}(\alpha_s)$ has been given in (\ref{gcusp2}), and 
\begin{equation}
   \gamma_{\rm cusp}^{(2)}(\beta_t) 
   = \frac{N\alpha_s^2}{2\pi^2} 
    \left[ \frac{i\pi}{2\beta_t} \left( 2 - \frac{\pi^2}{6} \right) 
    - 1 + \zeta_3 \right]  
\end{equation}
arises from the threshold expansion of the two-loop coefficient of the recoil-dependent cusp anomalous dimension in (\ref{Kido}). 

We stress the important fact that, as a consequence of the Coulomb singularities present in the function $g(\beta)$, the three-parton correlation terms governed by $f_2$ do {\em not\/} vanish near threshold. Instead, they give rise to scattering-angle dependent, off-diagonal contributions in (\ref{Gqqthresh}) and (\ref{Gggthresh}). These off-diagonal terms were not considered in two recent papers \cite{Beneke:2009rj,Czakon:2009zw}, where threshold resummation for top-quark pair production was studied at NNLL order. Explicit results equivalent to the threshold-expanded anomalous-dimension matrices (\ref{Gqqthresh}) and (\ref{Gggthresh}) were given in \cite{Czakon:2009zw}. We agree with those expressions up to the terms originating from $f_2$. We leave it to future work to explore if and how the results obtained by these authors need to be modified in light of our findings.

\subsection{IR poles in the $\bm{q\bar q\to t\bar t}$ and $\bm{gg\to t\bar t}$ scattering amplitudes}
\label{subsec:IRpoles}

The results (\ref{eq:qqmatrix}) and (\ref{eq:ggmatrix}) for the anomalous-dimension matrices allow us to construct the IR divergences in the virtual corrections to the partonic amplitudes for top-quark pair production at two-loop order. The key relation is that the IR poles in the partonic scattering amplitudes can be absorbed into a multiplicative renormalization factor, as shown in (\ref{Mfinite}). Expanding this relation in powers of $\alpha_s/4\pi$ yields
\begin{equation}\label{eq:IRamps}
\begin{split}
   |{\cal M}_n^{(1),\,\rm sing}\rangle 
   &= \bm{Z}^{(1)}\,|{\cal M}_n^{(0)}\rangle \,, \\
   |{\cal M}_n^{(2),\,\rm sing}\rangle 
   &= \left[ \bm{Z}^{(2)} - \left( \bm{Z}^{(1)}\right)^2 \right]
    |{\cal M}_n^{(0)}\rangle
    + \left( \bm{Z}^{(1)}\,|{\cal M}_n^{(1)}\rangle \right)_{\rm poles} ,
\end{split}
\end{equation}
where we have written the UV-renormalized QCD amplitudes as
\begin{equation}
   |{\cal M}_n\rangle = 4\pi\alpha_s \left[ |{\cal M}_n^{(0)}\rangle 
    + \frac{\alpha_s}{4\pi}\,|{\cal M}_n^{(1)}\rangle 
    + \left( \frac{\alpha_s}{4\pi} \right)^2 |{\cal M}_n^{(2)}\rangle 
    + \dots \right] .
\end{equation}
Note that to predict the IR poles at two-loop order, one needs the UV-renormalized one-loop amplitudes to ${\cal O}(\epsilon)$, since these terms combine with the $1/\epsilon^2$ piece of the one-loop $\bm{Z}$ factor to give a $1/\epsilon$ pole term.

In evaluating the relations (\ref{eq:IRamps}), one must make sure that the UV-renormalized amplitudes and the $\bm{Z}$ factors are expressed in terms of one and the same coupling constant. As mentioned in Section~\ref{sec:ADs}, one possibility is to express all results in terms of $\alpha_s$ defined with five active flavors, by applying the decoupling relation (\ref{eq:decouple}) to the UV-renormalized QCD amplitudes. After doing so, any dependence on the number $n_h$ of heavy-quark flavors drops out of the results for the pole terms. Alternatively, by applying the inverse decoupling relation to the coupling $\alpha_s$ in the $\bm{Z}$ factors, one can express the perturbative series in terms of $\alpha_s^{\rm QCD}$ defined in full QCD with six active flavors ($n_h=1$ for the top quark, and $n_l=5$ for the remaining quarks). One then recovers the full $n_h$ dependence in the singular parts of the QCD amplitudes. At the level of the renormalization factor, the conversion to six active flavors is accomplished by adding an additional piece to the expression given in \cite{Becher:2009cu,Becher:2009qa}. At two-loop order, we find the general result
\begin{equation}\label{result}
\begin{split}
   \bm{Z} &= 1 + \frac{\alpha_s^{\rm QCD}}{4\pi} 
    \left( \frac{\Gamma_0'}{4\epsilon^2}
    + \frac{\bm{\Gamma}_0}{2\epsilon} \right) \\
   &\quad\mbox{}+ \left( \frac{\alpha_s^{\rm QCD}}{4\pi} \right)^2 
    \Bigg\{ \frac{(\Gamma_0')^2}{32\epsilon^4} 
    + \frac{\Gamma_0'}{8\epsilon^3} 
    \left( \bm{\Gamma}_0 - \frac32\,\beta_0 \right) 
    + \frac{\bm{\Gamma}_0}{8\epsilon^2} 
    \left( \bm{\Gamma}_0 -2\beta_0 \right) 
    + \frac{\Gamma_1'}{16\epsilon^2}
    + \frac{\bm{\Gamma}_1}{4\epsilon} \\
   &\qquad\mbox{}- \frac{2T_F}{3}\,\sum_{i=1}^{n_h}\, 
    \Bigg[ \Gamma_0' \left( \frac{1}{2\epsilon^2}\,\ln\frac{\mu^2}{m_i^2}
    + \frac{1}{4\epsilon} \left[ \ln^2\!\frac{\mu^2}{m_i^2}     
    + \frac{\pi^2}{6} \right] \right)
    + \frac{\bm{\Gamma}_0}{\epsilon}\,\ln\frac{\mu^2}{m_i^2} \Bigg]
    \Bigg\} + {\cal O}(\alpha_s^3) \,.
\end{split}
\end{equation}
The coefficients $\bm{\Gamma}_n$ are defined via the expansion 
\begin{equation}
   \bm{\Gamma} = \sum_{n\ge 0}\,\bm{\Gamma}_n 
   \left( \frac{\alpha_s}{4\pi} \right)^{n+1} ,
\end{equation}
and similarly for the quantity $\Gamma'=-2C_i\,\gamma_{\rm cusp}(\alpha_s)$, where $C_i=C_F$ for the $q\bar q$ channel, and $C_i=C_A$ for the $gg$ channel. The $f_2$ term enters the two-loop $1/\epsilon$ pole via $\bm{\Gamma}_1/\epsilon$ in (\ref{result}). We emphasize that in the $\beta$-function coefficient $\beta_0=\frac{11}{3}\,C_A-\frac43\,T_F n_l$ and in the two-loop anomalous-dimension coefficients $\bm{\Gamma}_1$ and $\Gamma_1'$ in (\ref{result}) the number $n_l$ of active flavors only includes the massless quarks, not the massive ones. The $n_h$ dependence of the full-theory $\bm{Z}$ factor is contained entirely in the terms shown in the third line.

The result (\ref{eq:IRamps}) is an exact prediction for the IR poles of the partonic amplitudes at two-loop order, which can be tested against explicit loop calculations. In practice, however, one is interested mainly in the real part of the interference terms $\langle{\cal M}_n^{(0)}|{\cal M}_n^{(2)}\rangle$, since it is these which are needed to calculate partonic cross sections. For this reason, we give results for the interference terms rather than the amplitudes themselves. For the specific case of $t\bar t$ production, the full results for both the $q\bar q$ and $gg$ channels are rather lengthy and are included as a computer program in the electronic version of this paper. In what follows, we will define the color decomposition used at two-loop order and describe to what extent we can compare our results with those available in the literature. As explained below, the three-parton correlations proportional to $f_2$ do not appear in the interference of the Born level and two-loop amplitudes, neither in the $q\bar q$ channel nor in the $gg$ channel.

For the $q\bar q\to t\bar t$ channel, the result for the interference term between the Born and two-loop amplitudes can be decomposed into color structures according to \cite{Czakon:2008zk}
\begin{equation}
\begin{split}
   2\,{\rm Re}\,\langle{\cal M}^{(0)}|{\cal M}^{(2)}\rangle_{q\bar q} 
   &= 2(N^2-1)\,\bigg( N^2 A^q + B^q + \frac{1}{N^2}\,C^q 
    + N n_l\,D_l^q + N n_h\,D_h^q \\
   &\hspace{22mm}\quad\mbox{}+ \frac{n_l}{N}\,E_l^q 
    + \frac{n_h}{N}\,E_h^q + n_l^2 F_l^q + n_l n_h\,F_{lh}^q 
    + n_h^2 F_h^q \bigg) \,.
\end{split}
\end{equation}
To compute the IR poles in the color coefficients above, we evaluate the general relation (\ref{eq:IRamps}) using (\ref{result}) for the renormalization factor and (\ref{eq:qqmatrix}) for the anomalous-dimension matrix. In addition, we need the finite parts of the UV-renormalized one-loop QCD amplitude up to ${\cal O}(\epsilon)$, decomposed into the singlet-octet color basis. We have obtained these through direct calculation, using some of the master integrals computed in \cite{Korner:2004rr}. After enforcing momentum conservation, the color coefficients are functions of the invariants $s=s_{12}$, $t_1=s_{13}=s_{24}$, $m_t$, and $\mu$. We have verified that our results for the IR poles agree with the numerical ones from \cite{Czakon:2008zk}, with the analytic results for some of the color coefficients given in \cite{Bonciani:2008az,Bonciani:2009nb}, and with the results in the small-mass limit from \cite{Czakon:2007ej}. In our case, all pole coefficients are available in analytic form. Since the Born-level $q\bar q\to t\bar t$ amplitude is proportional to the color-octet structure in (\ref{eq:qqbasis}) and the three-parton correlations proportional to $f_2$ enter the anomalous-dimension matrix (\ref{eq:qqmatrix}) only in the off-diagonal terms, the contributions from $f_2$ in the squared matrix element first appears at three-loop order. This was noted independently in \cite{Czakon:2009zw}. 

For the $gg\to t\bar t$ channel, we follow \cite{Czakon:2007wk} and decompose the interference term between the Born and two-loop amplitudes into color structures as
\begin{equation}\label{eq:colordecomp}
\begin{split}
   2\,{\rm Re}\,\langle {\cal M}^{(0)}|{\cal M}^{(2)}\rangle_{gg} 
   &= (N^2-1)\,\bigg( N^3 A^g + N\,B^g + \frac{1}{N}\,C^g 
    + \frac{1}{N^3}\,D^g \\
   &\quad\mbox{}+ N^2 n_l\,E_l^g + N^2 n_h\,E_h^g 
    + n_l\,F_l^g + n_h\,F_h^g + \frac{n_l}{N^2}\,G_l^g 
    + \frac{n_h}{N^2}\,G_h^g \\
   &\quad\mbox{}+ N n_l^2 H_l^g + N n_l n_h\,H_{lh}^g 
    + N n_h^2 H_h^g + \frac{n_l^2}{N}\,I_l^g 
    + \frac{n_l n_h}{N}\,I_{lh}^g + \frac{n_h^2}{N}\,I_h^g \bigg) \,.
\end{split}
\end{equation}
The IR poles in the color coefficients are obtained as for the $q\bar q$ channel, except in this case we use the anomalous-dimension matrix (\ref{eq:ggmatrix}). Results in the literature are available only in the small-mass limit \cite{Czakon:2007wk}, and we have checked the agreement of our exact results with this limiting case. Since the exact results are new, we list in Table \ref{tab:GGpoles} the numerical values for the poles of the color coefficients at the point $t_1=-0.45s$, $s=5m_t^2$, and $\mu=m_t$. Again in this case the results do not depend on $f_2$, the reason being that the corresponding contribution is multiplied by a color structure which is anti-symmetric under the exchange of the two initial-state gluons, while the $gg\to t\bar t$ amplitude is symmetric under this exchange.

\begin{table}[t]
\begin{center}
\begin{tabular}{||c|r|r|r|r||}
\hline\hline
 & $\epsilon^{-4}\quad$ & $\epsilon^{-3}\quad$ & $\epsilon^{-2}\quad$
 & $\epsilon^{-1}\quad$ \\
\hline\hline
$A^g$ & 10.749 & 18.694 & $-156.82$ & 262.15 \\
$B^g$ & $-21.286$ & $-55.990$ & $-235.04$ & 1459.8 \\
$C^g$ & & $-6.1991$ & $-68.703$ & $-268.11$ \\
$D^g$ & & & 94.087 & $-130.96$ \\
$E_l^g$ & & $-12.541$ & 18.207 & 27.957 \\
$E_h^g$ & & & 0.012908 & 11.793 \\
$F_l^g$ & & 24.834 & $-26.609$ & $-50.754$ \\
$F_h^g$ & & & 0.0 & $-23.329$ \\
$G_l^g$ & & & 3.0995 & 67.043 \\
$G_h^g$ & & & & 0.0 \\
$H_l^g$ & & & 2.3888 & $-5.4520$ \\
$H_{lh}^g$ & & & & $-0.0043025$ \\
$H_h^g$ & & & & \\
$I_l^g$ & & & $-4.7302$ & 10.810 \\
$I_{lh}^g$ & & & & 0.0 \\
$I_{h}^g$ & & & & \\ 
\hline\hline
\end{tabular}
\parbox{15cm}{\caption{\label{tab:GGpoles} 
Numerical results for the IR poles in the color coefficients (\ref{eq:colordecomp}) for top-quark pair production in the $gg\to t\bar t$ channel, evaluated at the point $t_1=-0.45 s$, $s=5m_t^2$, and $\mu=m_t$. The blank entries are not present in general, while the entries with value 0.0 vanish only for the particular choice $\mu=m_t$.}}
\end{center}
\end{table}

\section{Elastic quark-quark scattering in the forward limit}
\label{sec:elqq}

Another interesting application of our general formalism is the case of elastic quark-quark scattering at high energy and fixed momentum transfer ($s,m^2\gg|t|$). The anomalous-dimension matrix for this case was analyzed at two-loop order in \cite{Korchemskaya:1994qp} by studying the cross singularities of self-intersecting Wilson loops. We will now show that the results derived in that paper can be obtained by taking a certain limit of our general results, and that this provides a cross-check on our calculation of the three-parton correlations governed by the function $F_1$.   

Consider the elastic process $q_{1j}(p_1)+q_{2l}(p_2)\to q_{1i}(p_3)+q_{2k}(p_4)$ for massive quarks ($m_1=m_2\equiv m$) in the forward limit
\begin{equation}
  s,\,m^2\gg -t\gg\Lambda_{\rm QCD}^2 \,.
\end{equation}
Here $i,j,k,l$ are color indices, and 1,2 label the quark flavors. The relevant cusp angles can be expressed in terms of the invariants $s=(p_1+p_2)^2$ and $t=(p_1-p_3)^2$ as
\begin{equation}
\begin{aligned}
   \beta_{12} &= \beta_{34} 
   = \mbox{arccosh}\Big( - \frac{s-2m^2}{2m^2} \Big)
   \equiv i\pi - \gamma \,, \\
   \beta_{13} &= \beta_{24} 
   = \mbox{arccosh}\Big( \frac{2m^2-t}{2m^2} \Big)
   = {\cal O}\Big(\frac{\sqrt{-t}}{m}\Big) \,, \\
   \beta_{14} &= \beta_{23} 
   = \mbox{arccosh}\Big( \frac{s+t-2m^2}{2m^2} \Big) = \gamma +
   {\cal O}\Big(\frac{t}{m^2}\Big) \,,
\end{aligned}
\end{equation}
where $\cosh\gamma=v_1\cdot v_2=p_1\cdot p_2/m^2$. In the limit $t/m^2\to 0$ these angles are described in terms of a single variable $\gamma>0$.  Starting from the general expression (\ref{resu1}), we then obtain for the cross anomalous-dimension matrix
\begin{equation}\label{eq:gQQe}
\begin{split}
   \bm{\Gamma}_{\rm cross}(\gamma,\alpha_s)
   &\equiv \bm{\Gamma}_{qq}(s,t,m^2,\mu) \big|_{-t\ll s,m^2} \\
   &= - 2 \Big[ 
    \bm{T}_1\cdot\bm{T}_2\,\gamma_{\rm cusp}(i\pi-\gamma,\alpha_s) 
    + \bm{T}_1\cdot\bm{T}_3\,\gamma_{\rm cusp}(0,\alpha_s)
    + \bm{T}_1\cdot\bm{T}_4\,\gamma_{\rm cusp}(\gamma,\alpha_s) \Big] \\
   &\quad\mbox{}+ 4\gamma^Q(\alpha_s) 
    + 24if^{abc}\,\bm{T}_1^a \bm{T}_2^b \bm{T}_3^c\,
    F_1(i\pi-\gamma,\gamma,0) + {\cal O}(\alpha_s^3) \,,
\end{split}
\end{equation}
where we have used color conservation to simplify the various terms. The velocity-dependent cusp anomalous dimension $\gamma_{\rm cusp}(\beta,\alpha_s)$ has been given in (\ref{Kido}). Moreover, noting that the case at hand corresponds to the zero-recoil limit discussed in Section~\ref{sec:prop}, we can read off $F_1$ from (\ref{eq:zrf1f2}):
\begin{equation}
   F_1(i\pi-\gamma,\gamma,0)
   = \frac{\alpha_s^2}{12\pi^2}\,\Big[ \pi^2 A(\gamma)
    + i\pi\,B(\gamma) \Big] \,,
\end{equation}
where $A$ and $B$ were defined in (\ref{eq:ab}).

To give explicit results for the anomalous-dimension matrix (\ref{eq:gQQe}), we must first specify a color basis. We shall use the $t$-channel singlet-octet basis, where the two color structures are
\begin{equation}
   |c_1\rangle = \delta_{ij}\,\delta_{kl} \,, \qquad 
   |c_2\rangle = t^a_{ij}\,t^a_{kl} \,.
\end{equation}
In this basis, the result for the cross anomalous dimension valid to two-loop order is
\begin{equation}\label{eq:GammaCross}
\begin{aligned}
   \Big[ \bm{\Gamma}_{\rm cross}(\gamma) \Big]_{11} &= 0 \,, \\
   \Big[ \bm{\Gamma}_{\rm cross}(\gamma) \Big]_{12} 
   &= \frac{C_F}{N}\,\Big[ \gamma_{\rm cusp}(\gamma,\alpha_s)
    - \gamma_{\rm cusp}(i\pi-\gamma,\alpha_s) \Big]
    + 6 C_F F_1(i\pi-\gamma,\gamma,0) \,, \\
   \Big[ \bm{\Gamma}_{\rm cross}(\gamma) \Big]_{21} 
   &= 2\,\Big[ \gamma_{\rm cusp}(\gamma,\alpha_s)
    - \gamma_{\rm cusp}(i\pi-\gamma,\alpha_s) \Big]
    - 12 N F_1(i\pi-\gamma,\gamma,0) \,, \\
   \Big[ \bm{\Gamma}_{\rm cross}^{(2)}(\gamma) \Big]_{22} 
   &= \frac{2}{N}\,\Big[ \gamma_{\rm cusp}(i\pi-\gamma,\alpha_s)
    - \gamma_{\rm cusp}(\gamma,\alpha_s) \big]
    + N\,\Big[ \gamma_{\rm cusp}(\gamma,\alpha_s)
    - \gamma_{\rm cusp}(0,\alpha_s) \Big] \,.
\end{aligned}
\end{equation}
We have simplified the diagonal matrix elements using that $4\gamma^Q(\alpha_s)=-2C_F\gamma_{\rm cusp}(0,\alpha_s)$, which follows from the expression for the anomalous dimension of a heavy-quark current derived in \cite{Becher:2009kw}. Expanding the cross anomalous dimension as (this conforms with the notation used in~\cite{Korchemskaya:1994qp})
\begin{equation}
   \bm{\Gamma}_{\rm cross}(\gamma,\alpha_s)
   = \frac{\alpha_s}{\pi}\,\bm{\Gamma}_{\rm cross}^{(1)}(\gamma)
    + \left( \frac{\alpha_s}{\pi} \right)^2
    \bm{\Gamma}_{\rm cross}^{(2)}(\gamma) + \dots \,,
\end{equation}
the explicit one-loop result is
\begin{equation}
   \bm{\Gamma}_{\rm cross}^{(1)}(\gamma) 
   = \begin{pmatrix}
    0~ & ~\frac{C_F}{N}\,i\pi\coth\gamma \\
    2i\pi\coth\gamma~ & ~N(\gamma\coth\gamma-1) 
    - \frac{2i\pi}{N}\coth\gamma
   \end{pmatrix} ,
\end{equation}
whereas at two-loop order we obtain
\begin{eqnarray}
   \Big[ \bm{\Gamma}_{\rm cross}^{(2)}(\gamma) \Big]_{11} &=& 0 \,, 
    \nonumber\\
   \Big[ \bm{\Gamma}_{\rm cross}^{(2)}(\gamma) \Big]_{12} 
   &=& C_F \left[ \pi^2 A(\gamma) + i\pi \left( B(\gamma) 
    + \left[ \frac{31}{36} - \frac{5 T_F}{9N}\,n_l \right] \coth\gamma 
    \right) \right] , \nonumber\\
   \Big[ \bm{\Gamma}_{\rm cross}^{(2)}(\gamma) \Big]_{21} 
   &=& N \left( \frac{31}{18} - \frac{10 T_F}{9N}\,n_l \right)
    i\pi\coth\gamma \,, \nonumber\\[1cm]
   \Big[ \bm{\Gamma}_{\rm cross}^{(2)}(\gamma) \Big]_{22} 
   &=& \frac{N^2}{2} \Bigg\{
    \coth^2\gamma \left[ \mbox{Li}_3(e^{-2\gamma}) 
    + \gamma\,\mbox{Li}_2(e^{-2\gamma}) - \zeta_3 
    + \frac{\pi^2}{6}\,\gamma + \frac{\gamma^3}{3} \right] \\
   &&\mbox{}+ \coth\gamma \left[ \mbox{Li}_2(e^{-2\gamma}) 
    - 2\gamma\ln(1-e^{-2\gamma}) - \frac{\pi^2}{6} 
    + \left( \frac{67}{18} - \frac{\pi^2}{3} \right) \gamma - \gamma^2 
    - \frac{\gamma^3}{3} \right] \nonumber\\
   &&\mbox{}+ \gamma^2 + \frac{\pi^2}{3} - \frac{49}{18} \Bigg\} 
    - \frac59\,N T_F n_l \left( \gamma\coth\gamma - 1 \right) \nonumber\\
   &&\mbox{}- \pi^2 A(\gamma) - i\pi \left( B(\gamma) 
    + \left[ \frac{31}{18} - \frac{10 T_F}{9N}\,n_l \right] 
    \coth\gamma \right) . \nonumber
\end{eqnarray}
In deriving these expressions we have used the remarkable relation
\begin{equation}
\begin{split}
   \gamma_{\rm cusp}(\gamma,\alpha_s)
    - \gamma_{\rm cusp}(i\pi-\gamma,\alpha_s)
   &= i\pi\coth\gamma \left[ \frac{\alpha_s}{\pi}
    + \left( \frac{\alpha_s}{\pi} \right)^2 \left( \frac{31}{36}\,N
    - \frac59\,T_F n_l \right) \right] \\
   &\quad\mbox{}+ 6N F_1(i\pi-\gamma,\gamma,0)
    + {\cal O}(\alpha_s^3) \,,
\end{split}
\end{equation}
which links the complicated $\gamma$-dependent terms in the difference of cusp anomalous dimensions to those in $F_1(i\pi-\gamma,\gamma,0)$. In order to compare with the expressions given in \cite{Korchemskaya:1994qp}, we must convert our results to the color basis
\begin{equation}
   |c'_1\rangle = \delta_{ij}\,\delta_{kl} \,, \qquad 
   |c'_2\rangle = \delta_{il}\,\delta_{kj} \,.
\end{equation}
Results in that basis can be obtained from ours by the rotation
\begin{equation}
   \bm{\Gamma}'_{\rm cross}(\gamma,\alpha_s) 
   = \bm{V}\,\bm{\Gamma}_{\rm cross}(\gamma,\alpha_s)\,\bm{V}^{-1} ,
\end{equation}
where
\begin{equation}
   \bm{V} = \begin{pmatrix} 1~ & -\frac{1}{2N} \\
    0~ & ~\frac{1}{2} \end{pmatrix} , \qquad
   \bm{V}^{-1} = \begin{pmatrix} 1~ & ~\frac{1}{N} \\
    0~ & ~2 \end{pmatrix} .
\end{equation}
After performing this conversion, we find complete agreement with the results in \cite{Korchemskaya:1994qp}.\footnote{The cross anomalous dimension calculated here is the transpose of that in \cite{Korchemskaya:1994qp}, due to a different ordering of the matrices in the renormalization-group equation (\ref{eq:rgeo}) compared to Eq.~(2.9) in that paper. Also, these authors set $n_l=0$.} 
Since at two-loop order the off-diagonal elements in (\ref{eq:GammaCross}) receive contributions from $F_1$, this is a non-trivial check on our general result for this function.

\section{Conclusions}
\label{sec:concl}

The IR divergences of scattering amplitudes in non-abelian gauge theories can be absorbed into a multiplicative renormalization factor, whose form is determined by an anomalous-dimension matrix in color space. At two-loop order this anomalous-dimension matrix contains pieces related to color and momentum correlations between three partons, as long as at least two of them are massive. This information is encoded in two universal functions: $F_1$, describing correlations between three massive partons, and $f_2$, describing correlations between two massive and one massless parton. In this paper we have calculated these functions at  two-loop order. For $F_1$, this involved extracting the UV divergences of the vacuum matrix element of an HQET operator built out of three soft Wilson lines. The most complicated technical aspect of the calculation was the evaluation of a non-planar two-loop diagram, which we accomplished using Mellin-Barnes representations. The function $f_2$ was then obtained from $F_1$ by taking the limit where one of the three partons becomes massless.   

Using the exact analytic expressions, we studied the properties of the three-parton correlations in the small-mass, zero-recoil, and threshold limits. We found that the functions $F_1$ and $f_2$ vanish as $(m_I m_J/s_{IJ})^2$ in the small-mass limit, in accordance with existing factorization theorems for massive scattering amplitudes \cite{Mitov:2006xs,Becher:2007cu}. On the other hand, and contrary to naive expectations, the two functions do not vanish in the threshold limit, where the velocities of two heavy partons become nearly equal. The reason is that Coulomb singularities arise in this limit, which compensate a zero resulting from the anti-symmetry under exchange of two velocity vectors. 

Our results allow for the calculation of the IR poles of an arbitrary on-shell, $n$-particle scattering amplitude at two-loop order, where any number of the $n$ partons can be massive. As an application, we have derived the anomalous-dimension matrices for top-quark pair production (in association with colorless particles such as electroweak or Higgs bosons) in the $q\bar q\to t\bar t$ and  $gg\to t\bar t$ channels. These matrices form the basis for soft-gluon resummation at NNLL order for general kinematics, and in particular near the production threshold. We will explore in future work to what extent the new off-diagonal entries in the anomalous-dimension matrices, arising from the three-parton correlation terms, affect the numerical results for $t\bar t$ production at the Tevatron and LHC. Finally, we have used these matrices to determine the IR poles in the virtual corrections to the $t\bar t$ production cross sections at two-loop order in closed analytic form. The corresponding expressions, which agree with those in the literature where they exist, are very lengthy and are provided in the form of a computer program.

As a second interesting application, we have studied the elastic quark-quark scattering amplitude in the forward limit, where $s,m^2\gg|t|$. In this special case, our general expression for the anomalous-dimension matrix reduces to the cross anomalous dimension studied a long time ago in \cite{Korchemskaya:1994qp}. Since the three-parton correlation terms in the anomalous dimension give a non-zero contribution in this example, the fact that we find full agreement with the two-loop expressions given in that paper provides a non-trivial check of our calculation. 

The present paper completes the study of IR divergences of two-loop scattering amplitudes with an arbitrary number of massive and massless external particles, and in arbitrary non-abelian (or abelian) gauge theories with massless gauge bosons. In spontaneously broken gauge theories, our results can still be used above the symmetry-breaking scale, where gauge-boson masses can be neglected at leading power. At the symmetry-breaking scale a matching is done onto a non-interacting theory, in which the massive gauge bosons are integrated out. In this way, it is possible to resum electroweak Sudakov logarithms using effective-theory methods, as worked out in detail in \cite{Chiu:2008vv}. The next step should now be to apply these general results to specific hadron-collider processes.

\vspace{3mm}
{\em Acknowledgments:\/}
We are grateful to Gregory Korchemsky for pointing out the relevance of \cite{Korchemskaya:1994qp}, which serves as a non-trivial check of our results. We would like to thank Thomas Becher, David Broadhurst, Johannes Henn, J\"urgen K\"orner, and Sven Moch for useful discussions. B.P.~is supported by the State of Rhineland-Palatinate via the Research Centre ``Elementary Forces and Mathematical Foundations''.

\end{document}